\documentclass[aps,showpacs,preprint]{revtex4}
\usepackage{epsfig,amssymb,bm,graphics}

\newcommand*{\fs}[1]{#1\!\!\!/}

\begin{document}


\title{
Photoproduction of $\bm \phi$  meson off deuteron near threshold}


 \author{A.I.~Titov$^{a,b}$ and B.~K\"ampfer$^{a,c}$}
 \affiliation{
 $^a$Forschungzentrum Dresden-Rossendorf, 01314 Dresden, Germany\\
 $^b$Bogoliubov Laboratory of Theoretical Physics, JINR,
  Dubna 141980, Russia\\
 $^c$ Institut f\"ur Theoretische Physik, TU~Dresden, 01062 Dresden,
 Germany
 }


\begin{abstract}
 We discuss coherent and incoherent
 $\phi$ meson photoproduction off the deuteron at low energy
 and small momentum transfer with the aim to check whether
 the recent experimental data need for their interpretation
 the inclusion of  exotic channels. Our analysis of the differential cross
 section and spin-density matrix elements shows that the existing data may be
 understood on the base of conventional dynamics.
 For a firm conclusion about a possible manifestation of exotic
 channels one has to improve the resolution of the data with
 providing additional information on channels with
 spin- and double-spin flip transitions being  sensitive
 to the properties of the photoproduction amplitude in $\gamma p$
 and $\gamma D$ reactions.
\end{abstract}


 \maketitle

\section{Introduction}

  The investigation of the $\phi$-meson photoproduction at
  low energies, $E_\gamma\simeq 1.6-3$ GeV, plays an important role in
  understanding the non-perturbative  Pomeron exchange
  dynamics and the nature of $\phi N$ interaction.
  It was expected
  that in the diffractive region
  the dominant contribution comes from the Pomeron exchange, since
  the processes associated with conventional meson (quark) exchanges
  are suppressed by the
  OZI rule~\cite{NakanoToki,PL97,TOYM98,ZLB98,Will98,Laget2000,TL03}.
  An example of such  a (suppressed) process is the
  pseudoscalar $\pi$ and $\eta$ meson exchange which,
  as a rule, were considered as a small correction to the dominant
  Pomeron exchange channel. The Pomeron exchange
  amplitude is usually described in terms of the
  Donnachie-Landshoff model~\cite{DL84-92}, where the Pomeron
  couples to  single constituent quarks as a $C=+1$ isoscalar
  photon or/and its two-gluon exchange
  modification~\cite{Laget2000,Ryskin93,Cudell97}.
  These models are designed for the vector meson
  photoproduction at high energy and small momentum transfer.
  The  validity of an extrapolation of these models
  into the low energy region  and close to the threshold is not clear.
  Near threshold, the models predict a monotonic increase of the
  differential cross section  of $\gamma p\to \phi p$
  reaction at forward photoproduction angle
  with energy. However, a recent analysis of the $\phi$
  photoproduction at low energy by the LEPS collaboration
  shows a sizeable deviation
  from this prediction, in particular, the data show a bump structure
  around $E_\gamma\simeq 2$~GeV~\cite{Mibe05}.
  Another peculiarity of the LEPS data is a strong deviation
  of the spin-density matrix element $\rho^1_{1-1}$ from 0.5,
  which is in favor of a sizable contribution of un-natural parity
  exchange processes. These facts rise several questions:
  (i) whether one has to modify the conventional Pomeron exchange
  model at low energy, (ii) what is the source of un-natural
  parity exchange channels, (iii) whether we need to introduce
  some exotic channels (additional Regge trajectories, processes
  associated with possible hidden strangeness in the nucleon, etc.)
  to describe the data. In principle, these questions are related
  to each other and have to be analyzed simultaneously. Thus, for
  example, the mentioned bump-like behavior may be a result of the interplay of
  the pseudoscalar exchange amplitude
  and modified Pomeron exchange channels.

   The coherent $\phi$ photoproduction off the deuteron in the diffraction region
   seems to be very useful for such an analysis. First of all,
   the isovector $\pi$-meson exchange amplitude is eliminated
   in case of the isoscalar target. Therefore, the appearance of the
   bump-like structure in the energy dependence of the
   differential cross section of the reaction $\gamma D\to\phi D$
   would favor a modification of the conventional Pomeron
   exchange amplitude. The next step is an analysis of spin
   observables, in particular, the properties of the
   decay $\phi\to K^+K^-$ with unpolarized and polarized photon
   beams.
   The incoherent $\phi$ photoproduction in $\gamma D\to\phi pn$
   reaction allows to extract
   observables of the reaction $\gamma n\to \phi n$ which can be used
   for a simultaneous
   analysis of photoproduction off  neutron and proton targets
   in order to get additional and independent hint to
   a manifestation of possible exotic channels.

\begin{figure}[ht]
   \includegraphics[width=0.3\columnwidth]
   {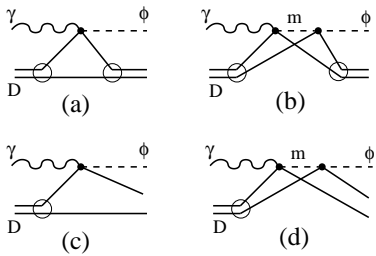}
   \caption{\small{Diagrammatic  representation of
   coherent (a,b) and in-coherent (c,d)
   $\phi$ meson photoproduction in $\gamma D$ reactions
   with single (a,c) and double (b,d) scattering contributions.}}
   \label{Fig:1}
   \end{figure}

   Schematically, the coherent and in-coherent $\phi$ meson
   photoproduction processes are exhibited in Fig.~\ref{Fig:1} (a,b) and
   (c,d), respectively.  The single and double
   scattering diagrams are shown in (a,b) and
  (c,d), respectively. The internal dashed line in (b) and (d)
  corresponds to "diagonal" ($m=\phi$) and "non-diagonal"
  ($m=\pi,\rho,\omega\,...$) transitions, respectively.
  In this paper we study the $\phi$ meson photoproduction
  at low energies with $E_\gamma<3$~GeV at forward
  photoproduction angles
  with momentum transfer $|t|\lesssim 0.4$~GeV$^2$, where the single
  scattering processes are dominant.
  The coherent $\phi$-meson
  photoproduction at higher values of $|t|$ is controlled by the double
  scattering processes, which can provide important information
  about the cross section of the $\phi N$ scattering~\cite{FKS97,Rogers06}.
  However, this interesting topic is beyond scope of our present
  analysis, where we focus just on the extremely forward $\phi$ meson
  photoproduction, where some hint to an "anomaly" in the differential
  cross section of $\gamma p\to \phi p$ reaction
  was found~\cite{Mibe05}. Some theoretical estimate for the coherent
  vector meson photoproduction from deuteron
  is given in Ref.~\cite{TFL02}. The first experimental data on
  $\gamma D\to \phi D$ reaction are reported recently in
  Refs.~\cite{Mibe07,WenChen07}.

    The aim of the present paper is to extend the results
    of Ref.~\cite{TFL02} for the coherent and incoherent
    $\phi$ meson photoproduction off the deuteron and give
    a consistent analysis of the recent experimental data
    towards understanding whether they can be described
    in terms of conventional dynamics or one needs to introduce
    some new (exotic) processes.

 Our paper is organized as follows.
 In Sec.~II we provide equations for the amplitudes of
 $\phi$ photoproduction off the proton which are used later on
 for coherent and incoherent $\phi$ meson photoproduction
 in $\gamma D$ reactions.
 Here we also analyze the unpolarized differential cross section
 of the reaction $\gamma p\to \phi p$.
  In Sec.~III we present a model of the coherent
  $\gamma D\to \phi D$ reaction.
 The incoherent $\gamma D\to \phi np$ reaction is considered in
 Sec.~IV. In Sec.~V we provide a simultaneous analysis of
 spin-density matrix elements for
 $\phi\to K^+K^-$ decay distributions in $\gamma p$,
 $\gamma n$, and $\gamma D$ reactions.
 The summary is given in Sec.~VI.

\section{$\bm\Phi$ meson photoproduction off the proton}

   For the reaction $\gamma p\to \phi p$,
   we define the kinematical variables with usual notation.
   The four-momenta of the
   incoming photon, outgoing vector meson, initial and final
   protons  are denoted as $k_\gamma$, $q_\phi$, $p$ and $p'$,
   respectively. The standard Mandelstam variables are
   defined as  $t = (p' - p)^2
   = (k_\gamma-q_\phi)^2$, $s \equiv W^2 = (p+k_\gamma)^2$.

In forward-angle photoproduction the $s$ and $u$ channels with an
intermediate nucleon and nucleon resonances are negligibly weak
and the main contribution comes from the Pomeron and pseudoscalar
($\pi,\eta$) meson exchange processes. The corresponding model for
the $\phi$ meson photoproduction in $\gamma p\to \phi p$ reaction
is described in Ref.~\cite{TL03}. However, for the sake of
completeness in this section we provide the main expressions for
the invariant amplitudes which will be used below.

 The photoproduction amplitude is expressed in  standard form
 \begin{eqnarray}
 T^{\gamma p\to\phi p}_{m_f\lambda_\phi;m_i\lambda_\gamma}
 =\bar u_f{\cal M}^{}_{\mu\nu}u_i\,
 {\varepsilon^*}^\mu_{\lambda_\phi}
 \varepsilon^\nu_{\lambda_\gamma}~,
 \label{P1}
 \end{eqnarray}
 where $\varepsilon_{\lambda_\gamma}$ and ${\varepsilon}_{\lambda_\phi}$
 are the polarization vectors of the photon and $\phi$ meson,
 respectively, and $u_i$=$u^{}_{m_i}(p)$
 [$u_f$=$u^{}_{m_f}(p')$]  is the
 Dirac spinor of the nucleon with momentum $p$ $[p']$ and spin projection
 $m_i$ $[m_f]$.

For the Pomeron exchange amplitude we utilize the modified
Donnachie-Landshoff~(DL) model~\cite{DL84-92}  to write
\begin{eqnarray}
 {\cal M}_{}^{\mu\nu}
 =M(s,t)\,\Gamma^{\mu\nu}~,
 \label{P2}
\end{eqnarray}
where the transition operator $\Gamma^{\mu\nu}$ reads
\begin{eqnarray}
\Gamma^{\mu\nu}=
 \fs{k}_\gamma(g^{\mu\nu}
 -\frac{q_\phi^\mu q_\phi^\nu}{q_\phi^2})
 -\gamma^\nu(k_\gamma^\mu -q_\phi^\mu\frac{k_\gamma\cdot q_\phi}{q_\phi^2} )
 -(q_\phi^\nu-\frac{\bar p^\nu k_\gamma\cdot q_\phi}
 {\bar p\cdot k_\gamma})(\gamma^{\mu} - \frac{\fs{q}_\phi
  q_\phi^\mu}{q_\phi^2})
 \label{P3}
\end{eqnarray}
with $\bar p=(p+p')/2$. The last term with $\bar p$ is added to
restore the gauge invariance~\cite{TL03}. The scalar function
$M_P(s,t)$ is described by the Regge parametrization,
 \begin{equation}
 M^{}_P (s,t)= C^{}_P \, F^{}_1(t) \, F^{}_2 (t)\,\frac{1}{s}
 \left(\frac{s}{s_P} \right)^{\alpha_P^{} (t)} \exp\left[ -
 \frac{i\pi}{2}\alpha_P^{}(t) \right]~,
 \label{P4}
 \end{equation}
where $F_1^{}(t)$ is the isoscalar  form factor of the nucleon and
$F_2^{}(t)$ is the form factor for the $\phi$
meson--photon--Pomeron coupling~\cite{DL84-92}
 \begin{eqnarray}
  F_1^{} (t) = \frac{ 4 M_N^2-a_N^2t }{ (4M_N^2-t) (1-t/t_0)^2  }~,\qquad
  F_2^{} (t) = \frac{2\mu_0^2}{(1 - t/M_\phi^2)(2 \mu_0^2 + M_\phi^2 -
  t)}~.
 \end{eqnarray}
 The Pomeron trajectory is known to be $\alpha_P^{} (t) = 1.08 +
0.25\,t$. The strength factor $C_P^{}$ is given by
\begin{eqnarray}
  C_P^{} = \frac{6eg^2}{\gamma_\phi^{}}~,
 \label{p5}
 \end{eqnarray}
 where $\gamma_\phi\simeq 6.7$ is the $\phi$ meson decay constant.
 The parameter $g^2$ is a product of two dimensionless coupling
 constants $g^2=g_{Pss}\cdot g_{Pqq}=
 (\sqrt{s_P}\beta_s)\cdot(\sqrt{s_P}\beta_u)$, where $g_{Pss}$
 and $g_{Pqq}$ have a meaning of the Pomeron coupling with the
 strange quark in  $\phi$ meson and light quark in a proton,
 respectively. In our study we choose:  $t_0 = 0.7$ GeV$^2$,
 $\mu_0^2=1.1$~GeV$^2$, $s_P=4$~GeV$^2$, $\beta_s=1.44$ and
 $\beta_{u(d)}=2.04$~GeV$^{-1}$. The parameter $a_N=2$ is taken to
 be larger than the corresponding parameter in
 DL model~\cite{DL84-92}, making
 the overall form factor close to that of the two-gluon exchange
 model~\cite{Cudell97}. Actually, the original DL model was
 motivated by the two-gluon exchange model of Landshoff and
 Nachtmann~\cite{Landshoff87}, therefore such a modification seems to
 be reasonable.

 In the case of the pseudoscalar
 mesons exchange ($M=\pi,\eta)$, the transition operator ${\cal M}^{}_{\mu\nu}$
 reads
\begin{eqnarray}
  {\cal M}_{\mu\nu}^{M}=
  -i\frac{eg_{\gamma\phi M}g_{M NN}}{M_\phi}\,\gamma_5\,
  \frac{\varepsilon^{\mu\nu\alpha\beta} {k_\gamma}_\alpha {{q_\phi}_\beta}}
  {t - M^2_\pi}
  F^2_M(t)
 \label{P6}
\end{eqnarray}
 with $g_{\pi NN}\simeq13.26 $,
 $g_{\gamma\phi\pi}\simeq-0.14$, and
 $g_{\gamma\phi\eta}\simeq-0.71$~\cite{TL03}. In this paper,
 following estimates based on QCD sum rule~\cite{QCDSR}
 and chiral perturbation theory~\cite{XPT}, as well as the
 phenomenological analysis of $\eta$
 photoproduction~\cite{etaphoto},
 we use $g_{\eta NN}\simeq 1.94$.
 $F_M^2$ is the product
of the two form factors of the virtual exchanged mesons in the
$MNN$ and $\gamma VM$ vertices
 \begin{eqnarray}
 F_M(t)=\frac{\Lambda_M^2-m_\pi^2}{\Lambda_M^22 - t}
\label{P7}
\end{eqnarray}
with $\Lambda_{\pi(\eta)}=1.05$~GeV. This value is slightly
greater than the values of cut-off parameters in Ref.~\cite{TL03}
($\Lambda_{\pi(\eta)}=0.6\,(0.9)$~GeV), which result in some
modification of the pseudoscalar exchange contribution. The SU(3)
symmetry predicts a constructive $\pi-\eta$ interference in
$\gamma p$ reactions and a destructive interference in $\gamma n$
reactions~\cite{TLTS99}.
    \begin{figure}[hb]
    \includegraphics[width=0.4\columnwidth]{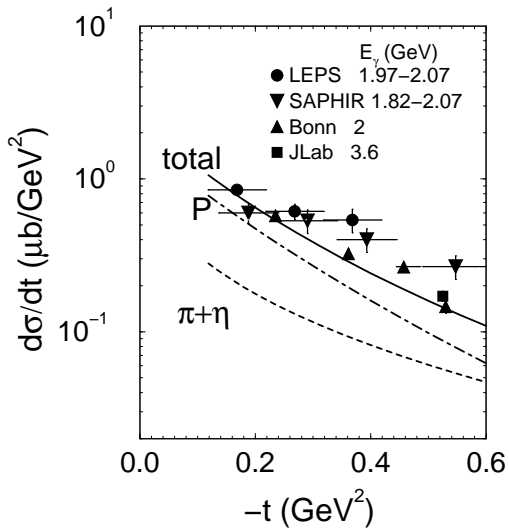}
   \caption{\small{Differential cross section of the $\gamma p\to \phi p$
   reaction as a function of momentum transfer $t$ at $E_\gamma=2.02$~GeV.
   The Pomeron, pseudoscalar exchange contributions and the
   total cross section
   are shown by dot-dashed, dashed and solid curves, respectively.
   Circles, triangles down, triangles up and squares  correspond to the
   LEPS~\protect~\cite{Mibe05},
   SAPHIR~\protect~\cite{SAPHIR},
  Bonn~\protect\cite{Bonn}, and
   JLab~\protect\cite{JLab36} data, respectively.
   \label{Fig:2}}}
   \end{figure}

 In Fig.~\ref{Fig:2} we show the differential cross section of the $\gamma p\to\phi p$
 reaction (solid curve) for the photon energy bin
 $E_\gamma=1.97-2.07$~GeV from LEPS~\cite{Mibe05}, together with the experimental
 data at $E_\gamma\sim2$~GeV~\cite{Mibe05,SAPHIR,Bonn}.
 For completeness, we also display
 JLab~\cite{JLab36} data, obtained at
 3.6~GeV, because there is no much difference in
 the $t$ dependence of Bonn~\cite{Bonn} and JLab~\cite{JLab36} data.
 One can see that the model  satisfactorily
 describes the Bonn and JLab experimental data. However, it underestimated
 the LEPS and SAPHIR data at relatively large $|t|$ which probably
 may manifest additional channels beyond our simple model~\cite{TL03}.
    \begin{figure}[hb!]
  \parbox{.45\textwidth}{
    \includegraphics[width=0.35\columnwidth]{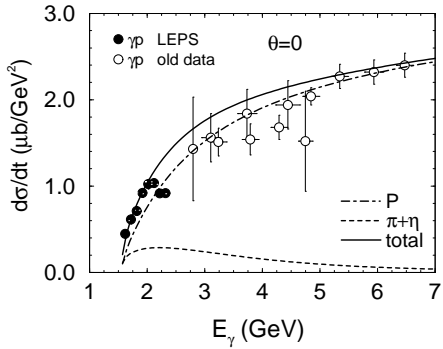}}
\hfill
  \parbox{.45\textwidth}{
   \caption{\small{Differential cross section of the $\gamma p\to \phi p$
   reaction at $t=t_{\rm max}$ ($\theta=0$)
   as a function of the photon energy.
   The experimental data are taken from~\protect\cite{Mibe05,OldData}.
    \label{Fig:3} }}}
   \end{figure}
 The energy dependence of the differential
 cross section at forward photoproduction angle with $\theta=0$
 (i.e. $t=t_{\rm max}$) together with the experimental
 data~\cite{Mibe05} is shown in  Fig.~\ref{Fig:3}.
  One can see a sizeable deviation of experimental data around
 $E_\gamma=2-2.3$~GeV from the monotonic theoretical curve, which
 is related to the difference between
$t$-dependence of different data sets
 and our model, discussed above.
 It is clear that for understanding the nature of this difference
 one needs more precise experimental data not only in differential
 cross section but in polarization observables sensitive to the
 spin flip channels at several energies.

\section{\boldmath $\bm \Phi$ meson photoproduction in $\gamma D$ reactions }

   In this section we consider coherent $\gamma D\to \phi D$ and
   incoherent $\gamma D\to \phi np$ photoproduction processes.
   The kinematical variables for these reactions
   are the following ones.
   The four-momenta of the
   initial and the final
   deuteron ($np$ system) are denoted as $p_D$ and $p'_{X}$ ($X=D,np$),
   respectively. The Mandelstam variables are
   defined as  $s_D \equiv W_D^2 = (p_D+k_\gamma)^2$,
   $t_{X} = (p'_{{X}} - p_{D})^2$, and so on.
   The space component of the momentum transfer to deuteron in
   the laboratory system is ${\bf q}^2\equiv q^2=-t_D(1-t_D/4M_D^2)$, where
   $M_D$ is the deuteron mass.

 \subsection{Coherent photoproduction}

 As mentioned above, here we consider the $\phi$ meson
 photoproduction at forward angles with $|t|\lesssim 0.4$~GeV$^2$, where the dominant
 contribution comes from the single scattering process, shown in
 Fig.~\ref{Fig:1}~{a}. In such a case one can use a non-relativistic
 framework for the deuteron form factor based on utilizing
 the realistic $NN$ interaction. In our analysis we use the deuteron
 wave function calculated with Paris potential~\cite{Paris,ParisD}
 designed just for describing nuclear processes at high momentum
 transfer. Thus, it describes fairly well the deuteron electromagnetic
 form factor with momentum transfer up to $-t\simeq
 0.9$~GeV$^2$~\cite{ParisD}.

The total vector meson photoproduction amplitude in the reaction
$\gamma
 D\to VD$ reads
 \begin{eqnarray}
 T^D_{M_fM_i;\lambda_V\lambda_\gamma}= 2 \sum_{\alpha\beta}
 \langle M_f\lambda_V,\beta|
 T^s_{\beta\alpha;\,\lambda_V\lambda_\gamma}
 |M_i\lambda_\gamma,\alpha\rangle,
\label{A1}
 \end{eqnarray}
 where $M_i,M_f,\lambda_\gamma$, and $\lambda_V$ stand for the  deuteron-spin
 projections of the initial and final states, and helicities of
 the incoming photon and the outgoing vector meson, respectively.
 $T^s$ is the amplitude of the vector meson
 photoproduction from the isoscalar nucleon
 \begin{eqnarray}
 T^s\equiv\frac12(T^p + T^n ).
 \label{A1_2}
 \end{eqnarray}
 The indices $\alpha$
 and $\beta$ in Eq.~(\ref{A1}) refer to all quantum numbers before and after
 the collision. The "elementary" photoproduction amplitudes
 $T^{p,n}$ are defined in the previous section.
 $\pi$ exchange terms are canceled  in the total amplitude since
 $T^n_\pi=-T^p_\pi$.

  Using the standard decomposition of the deuteron state
  in terms of $s$ $(U_0)$ and $d$ $(U_2)$ wave functions,
   one can rewrite Eq.~(\ref{A1}) in the explicit form
 \begin{eqnarray}
  T^D_{M_f,M_i;\lambda_V\lambda_\gamma}(t)&=&2\sqrt{4\pi}\sum
  i^\lambda\,\frac{\widehat{L'}\widehat{\lambda}}{\widehat{L}}
  Y_{\lambda\mu}(\widehat{\bf q})
\,
  C_{{\textstyle \frac12} m_1{\textstyle \frac12} m}^{1M}
  C_{{\textstyle \frac12} m_1'{\textstyle \frac12} m}^{1M'}
  C_{1M LM_L}^{1M_i}
  C_{1M' L'M_{L'}}^{1M_f}\nonumber\\
&&\qquad\times\, C_{L'M_{L'} \lambda\mu}^{LM_{L}}
  C_{L'0 \lambda  0}^{L0}
  \,R_{LL'\lambda}({q^2})\,
   T^s_{m_1m_1';\lambda_V\lambda_\gamma}(t),
   \label{A2}
 \end{eqnarray}
where $\widehat{j}=\sqrt{2j+1}$, and the radial  integral
$R_{LL'\lambda}$ reads
\begin{eqnarray}
  R_{LL'\lambda}(q^2)= \int dr
  U_L(r)U_{L'}(r)j_\lambda({qr}/{2}).
   \label{A3}
 \end{eqnarray}

For a qualitative analysis of the unpolarized differential cross
section at small momentum transfer with $\theta_{\hat{\bf
q}}\simeq0$,  keeping only the spin/helicity conserving terms with
natural $T^{\rm N}$ and unnatural $T^{\rm U}$ parity exchange in
the total amplitude, one gets
\begin{eqnarray}
  T^{{\rm N}\atop{\rm U}}_{mm';\lambda_V\lambda_\gamma}(t)=
  \left({{1}\atop{2m\lambda_\gamma}}\right)
  \delta_{mm'}\delta_{\lambda_\gamma\lambda_V}
  T^{{\rm N}\atop{\rm U}}_0(t).
   \label{A4}
 \end{eqnarray}
 Here, $T^{{\rm N}\atop{\rm U}}_0(t)$ is the spin-independent part of the
 amplitudes. Using
 Eq.~(\ref{A2}) with Eq.~(\ref{A4}), we get the following result for the
 natural and un-natural parity-exchange parts of the total  amplitude
\begin{eqnarray}
  &&T^{D{\rm N}}_{M_fM_i;\lambda_V\lambda_\gamma} =
  2\delta_{M_iM_f}\delta_{\lambda_\gamma\lambda_V}
  (\delta_{\pm1M_i}S_1^{\rm N}  
   +\delta_{0M_i}S_0^{\rm
  N})
   T^{{\rm N}}_0,\nonumber\\
&& T^{D{\rm U}}_{M_fM_i;\lambda_V\lambda_\gamma}=
   2M_i\lambda_\gamma\delta_{M_iM_f}\delta_{\lambda_\gamma\lambda_V}
  \delta_{\pm1M_i}\,S_1^{\rm U}\,
   T^{{\rm U}}_0.
   \label{A5}
 \end{eqnarray}
 The form factors $S_i^{N,U}$ read
 \begin{eqnarray}
  S^{N}_1&=& F_C - {\sqrt{2}}
  F_Q,\qquad
  S^{N}_0= F_C + 2\sqrt{2} F_Q,\qquad
  S^{U}_1= F_M~,
 \label{A6}
 \end{eqnarray}
 with
 \begin{eqnarray}
 F_C&=& R_{000} + R_{220},\qquad
 F_Q=R_{202} -\frac{1}{\sqrt{8}} R_{220}~,\nonumber\\
 F_M&=& R_{000} -\frac12 R_{220} + \sqrt{2} R_{202}+  R_{220}~.
 \label{A7}
 \end{eqnarray}
 Taking into account the cancelation of the un-natural parity $\pi$ exchange
 contribution and neglecting weak $\eta$ meson exchange,
 one can express the differential cross
 section of the $\gamma D\to \phi D$ reaction by
 the cross section of the $\phi$ photoproduction from the isoscalar
 nucleon $<N>$ as
\begin{eqnarray}
 \frac{d\sigma^{\gamma D}}{dt}\simeq
 \,4Z(t)\frac{d\sigma^{\gamma <N>}}{dt}~,
 \label{A8}
 \end{eqnarray}
 where $t=t_D$ and $Z(t)$ is the  structure factor
\begin{eqnarray}
 Z(t)=F_C^2(t)+4F_Q^2(t)~.
 \label{A8-1}
 \end{eqnarray}
    \begin{figure}[hb!]
  \parbox{.45\textwidth}{
    \includegraphics[width=0.35\columnwidth]{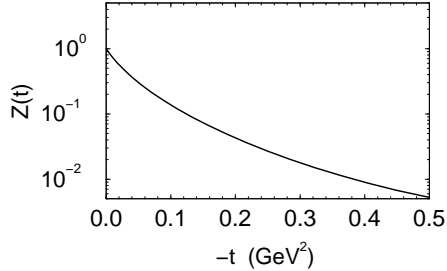}}
\hfill
  \parbox{.45\textwidth}{
   \caption{\small{The dependence of the structure factor $Z$
   on $t=t_D$.
    \label{Fig:4} }}}
   \end{figure}

 The dependence of $Z$ and $F_{C,Q}$ on $t_D$
 is rather symbolic. In fact,  these factors depend on
 the spatial part of the four-momentum transfer
 in the laboratory system $q$, as follows from Eq.~(\ref{A3}).
 The relation between $t_D$ and $q^2$ reads
 $t_D=-2M_D(\sqrt{q^2 +M_D^2}-M_D)$.
 The structure factor $Z$ as a function on $t_D$
 is shown in Fig.~\ref{Fig:4}. In the considered region of
 momentum transfer $t$, the factor $Z(t)$ is related to the
 well known structure function $A(t)$ of the elastic $eD\to eD$ scattering
 as
\begin{eqnarray}
 A(t)\simeq Z(t)\,G^2_d(t)~,
 \label{A8-1}
 \end{eqnarray}
where $G_d(t)=1/(1-t/0.71)^2$ is the dipole electromagnetic form
factor of the proton.

 Equation (\ref{A8}) allows to "extract" the cross section
 of the $\gamma <N>$ reaction from the measured cross section of the
 $\gamma D$ reaction as
\begin{eqnarray}
 \frac{d\sigma^{\gamma <N>}}{dt}\simeq
 \left[4Z(t)\right]^{-1}\frac{d\sigma^{\gamma D}}{dt}~.
 \label{A9}
 \end{eqnarray}

    \begin{figure}[hb!]
  \parbox{.45\textwidth}{
    \includegraphics[width=0.35\columnwidth]{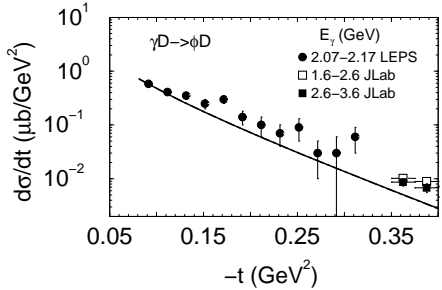}}
\hfill
  \parbox{.45\textwidth}{
   \caption{\small{Differential cross section of the $\gamma D\to \phi D$
   reaction  as a function of momentum transfer $t$ ($t=t_D$).
   Circles and squares correspond to
   LEPS~\protect~\cite{WenChen07}, and
   CLAS~\protect~\cite{Mibe07} data, respectively.
    \label{Fig:5} }}}
   \end{figure}

 In Fig.~\ref{Fig:5} the differential cross section
 of $\gamma D\to \phi D$ reaction is exhibited calculated by using
 the explicit expression for the photoproduction
 amplitude given by Eq.~(\ref{A1}), together with
 the available experimental data by LEPS
 (circles~\cite{WenChen07})
 and CLAS (squared~\cite{Mibe07})  collaborations.
 For simplicity, we show only a comparison for the bin
 $E_\gamma=2.07-2.17$~GeV.
 The description of the data for other bins has a similar quality.
 One can see that the model
 rather well describe the the data at low momentum transfers $|t_D|$
 but tends to underestimate the data at higher $|t|$, probably
 pointing to growing weight of more complicated (such as double scattering)
 channels.

\begin{figure}[hb!]
  \parbox{.45\textwidth}{
    \includegraphics[width=0.35\columnwidth]{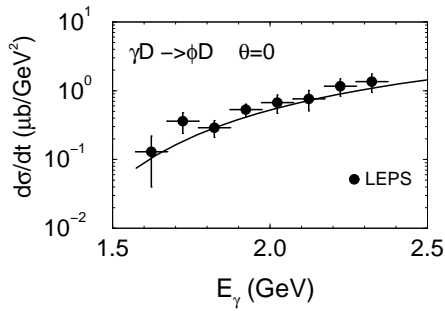}}
\hfill
  \parbox{.45\textwidth}{
   \caption{\small{Differential cross section of the $\gamma D\to \phi D$
   reaction at $t=t_{\rm max}$ ($t=t_D$)
   as a function of the photon energy.
   The experimental data are taken from~\cite{WenChen07}.
   \label{Fig:6}}}}
   \end{figure}

 In Fig.~\ref{Fig:6} we show the energy dependence of the differential
 cross section of the $\gamma D\to \phi D$ reaction
 at $\theta=0$ (i.e. $t=t_{\rm max}$)
 together with
 experimental data~\cite{WenChen07}. The agreement between data and
 model is fairly reasonable. Note that here the experimental data do not
 point to a bump-like structure at $E_\gamma\sim 2$~GeV.

\begin{figure}[tb!]
  \parbox{.45\textwidth}{
    \includegraphics[width=0.35\columnwidth]{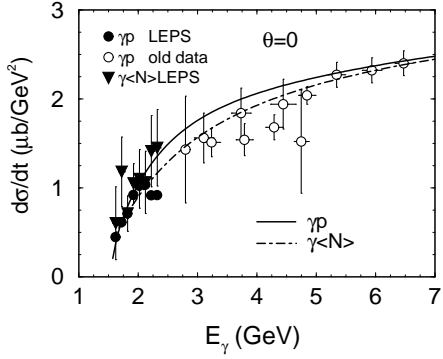}}
\hfill
  \parbox{.45\textwidth}{
   \caption{\small{Differential cross section of $\phi$ meson
   photoproduction off the proton (solid curve) and
   off the iso-scalar nucleon in $\gamma D$ reaction.
   The experimental data are taken from~\cite{Mibe05,WenChen07}.
   \label{Fig:7}}}}
   \end{figure}

In Fig.~\ref{Fig:7} the comparison of $\phi$ meson photoproduction
off the proton and off the isoscalar nucleon in a deuteron at
$\theta=0$ is displayed. In latter case the experimental data and
the theoretical curve are evaluated from the corresponding cross
section of the $\gamma D\to \phi D$ reaction by using
Eq.~(\ref{A9}). The figure displays the energy dependence of the
differential cross sections at $\theta=0$. One can see that the
two cross sections are close to each other at all energies. The
Pomeron exchange amplitude dominates ah high energies. At lower
energy, the behavior of cross sections of the $\gamma p$ and
$\gamma <N>$ reactions is not trivial. The elimination of the
isovector $\pi$
 exchange contribution in the $\gamma <N>$ reaction is compensated
 by a modification of momentum transfer $t$, which
 is smaller compared to that of $\gamma p$ reaction
 near the threshold in $\gamma D$ reaction.
 This causes the approach of both curves with decreasing energy
 $E_\gamma$.

\subsection{Incoherent photoproduction}

 The main purpose of the measurement and the theoretical
  study of the incoherent
 $\phi$-meson photoproduction in $\gamma D$ reactions is an extraction
 of the cross section
 of $\gamma n\to \phi n$ photoproduction with
 the goal of a subsequent combined analysis
 of  $\gamma p$ and $\gamma n$ reactions to seek for
 a possible manifestation of exotic channels.
 This problem seems not too difficult if one uses the
 exclusive $\gamma D\to \phi np$ reaction. But at low energy and forward
 photoproduction angles, the momenta of the recoil nucleons are
 small, and there is an experimental problem with their detection. Therefore,
 another way is to study the $[\gamma D,\phi]$ missing mass
 distribution in the inclusive $\gamma D\to\phi X$ ($X=np, D$)
 reaction. Below we develop a model which can be used for an extraction of
 the observables of $\gamma n\to \phi n$ photoproduction.

 The differential cross section of the $\phi$ meson photoproduction in the $\gamma
 D\to\phi np$ reaction reads
 \begin{eqnarray}
 \frac{d\sigma}{dt\,dM_X}
 =\frac{1}{16\pi(s-M_D^2)^2}\,
 \int d\widetilde{\Omega}\frac{\widetilde{p}}{16\pi^3}(|T_p|^2 +|T_n|^2)~,
 \label{B1}
 \end{eqnarray}
 where $\widetilde{p}$ and $\widetilde{\Omega}$ are the momentum
 and the solid angle of the spectator nucleon in the rest frame of
 the $np$ pair, respectively; $M_X$ is the invariant mass of this
 pair, and $t=t_X$; averaging and summing over the spin projections in the
 initial and the final states are assumed.
 $T_{p(n)}$ is the amplitude of the partial proton (neutron) contribution.
 It is related to the amplitude of the $\gamma N\to \phi N$ ($N=n,p$)
 reaction and the deuteron wave function $\psi^D$ as
 \begin{eqnarray}
 T_N=-\sqrt{2M_D}\sum\limits_{L\Lambda}
 \langle\frac12 m_2\frac12 \bar m|1M_i-\Lambda\rangle
 \langle L\Lambda 1M_i-\Lambda|1M_i\rangle\,
 T^{\gamma N\to \phi N}_{m_1\lambda_\phi;\bar m\lambda_\gamma}\,
 \psi^D_{L\Lambda}({\bf p}_s)
 \label{B2}
 \end{eqnarray}
 with
  \begin{eqnarray}
  \psi^D_{L\Lambda}({\bf p})&=&(2\pi)^{\frac32}i^L\,Y_{L\Lambda}(\widehat{\bf
  p})u_{L}(p),\nonumber\\
  u_{L}(p)&=&\sqrt{\frac{2}{\pi}}\int dr\,r\,U_L(r)\,j_{L}(pr)~,
 \label{B3}
  \end{eqnarray}
  where $p_s$ is the spectator momentum in the laboratory system,
  $u_L(r)$ is the radial deuteron wave function in the
  configuration space, $M_i$ $\lambda_\gamma$, $m_{1,2}$,
  and $\lambda_\phi$ are the spin
  projections of the incoming deuteron, photon helicity, the
  spin projections of the outgoing nucleons and the helicity
  of the $\phi$ meson, respectively.
    \begin{figure}[tb!]
           \parbox{.08\textwidth}{
       \includegraphics[width=0.04\columnwidth]{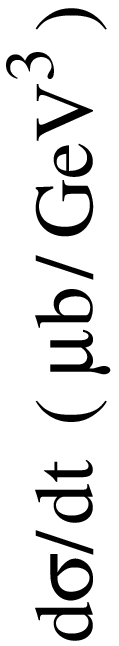}}
   \hspace{-3.5cm}
  \parbox{.85\textwidth}{
 \noindent
    \includegraphics[width=0.4527\columnwidth]{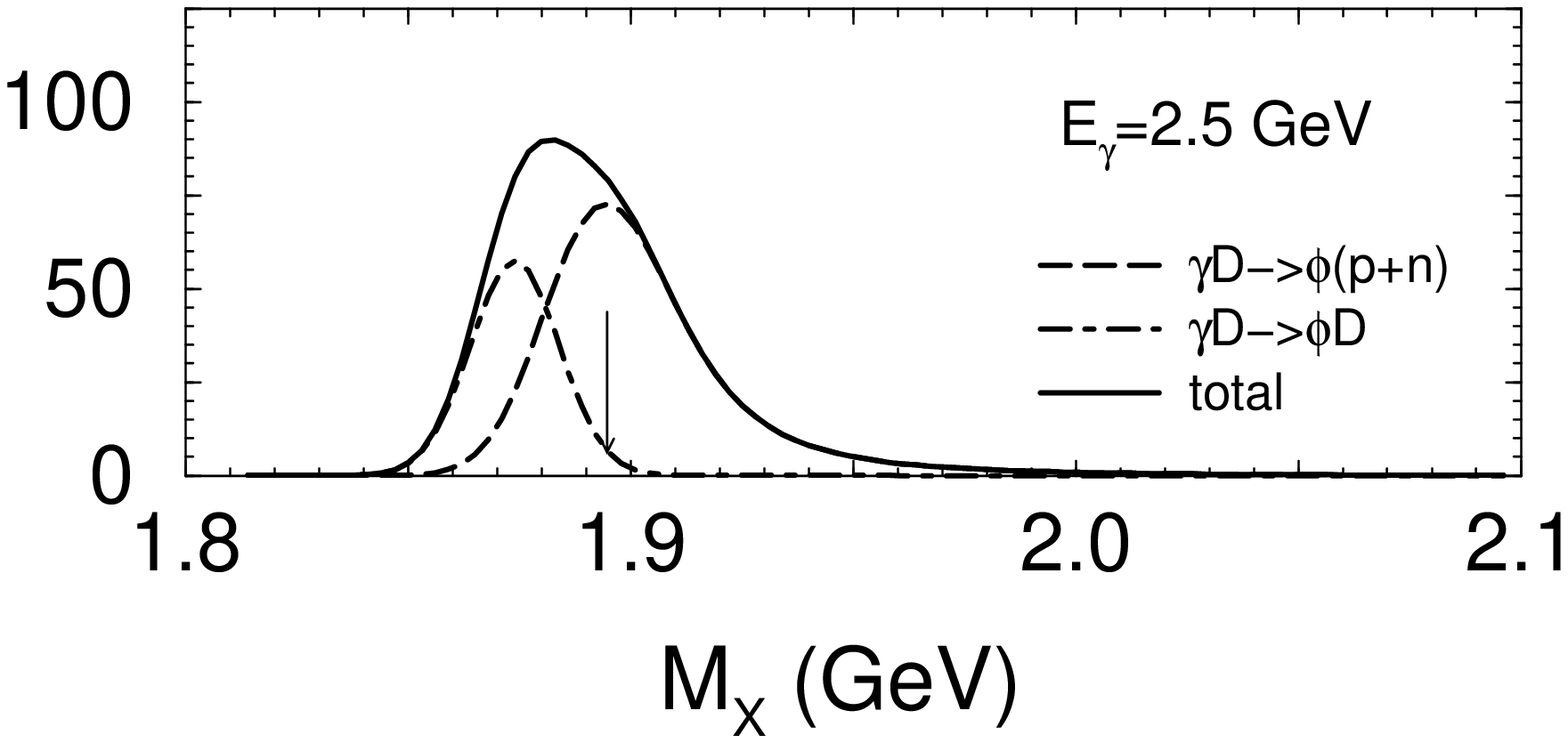}\\
  \vspace{-0.18cm}
\noindent
    \includegraphics[width=0.452\columnwidth]{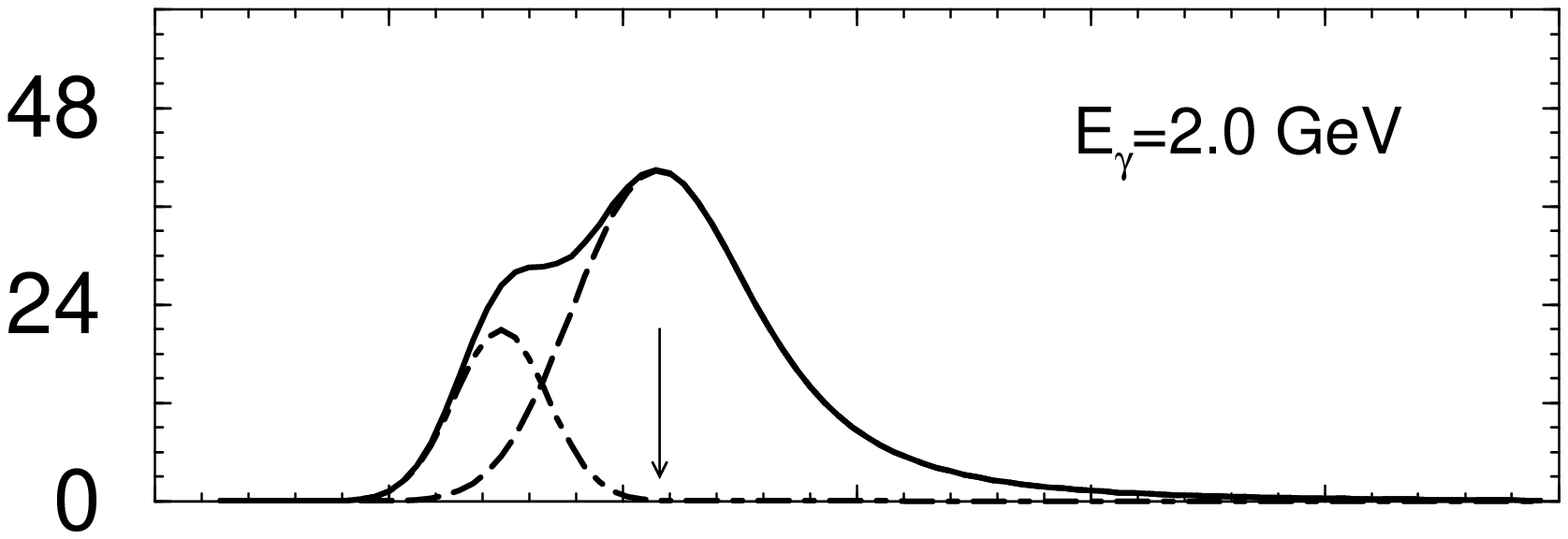}\\
  \vspace{-0.18cm}
    \includegraphics[width=0.475\columnwidth]{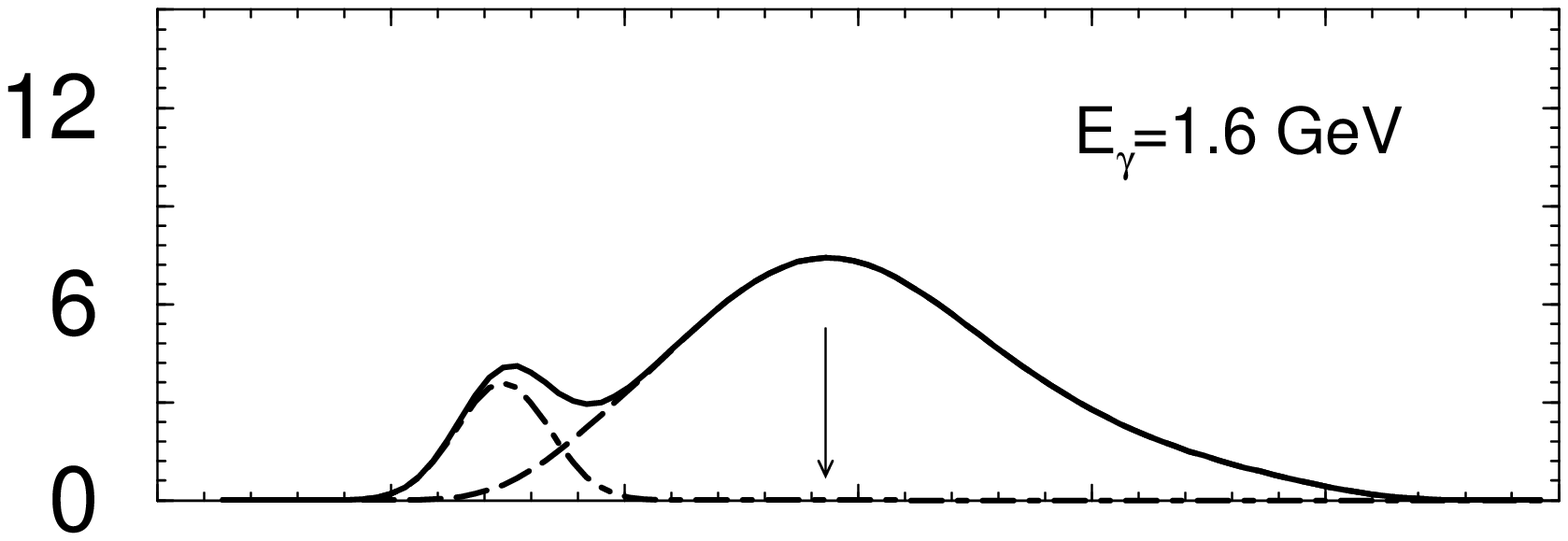}}
   \caption{\small{Differential distribution of $[\gamma D, \phi]$ missing
   mass in $\gamma D\to \phi X$ reactions at different energies.
   The curves  correspond to the cross sections
   of incoherent
   $\gamma D\to \phi (pn)$ (dashed) reactions,
   the coherent $\gamma D\to\gamma D$ (dot-dashed),
   and their sum (solid).
   The arrows mark the position of the maximum of the missing mass
   distribution.
   \label{Fig:8}}}
   \end{figure}
  For evaluating Eq.~(\ref{B1}) we define kinematical variables by the
  following steps. For given $M_X$, the energy of the outgoing
  nucleons in the $np$ rest frame is $\widetilde{E}=M_X/2$.
  Then, using $\widetilde{\Omega}$ and the $\phi$-meson
  photoproduction angle in the center of mass system
  as input variables we evaluate the four-momenta of the outgoing
  nucleons first in c.m.s. and then in the laboratory system.
  The four-momentum of the struck nucleon is  $p_i=p_D-p_s$,
  where $p_D=(M_D,{\bf 0})$. The amplitude $T^{\gamma N}$ in Eq.~(\ref{A2})
  is evaluated with an off-shell struck nucleon with $0<p_i^2<M_N^2$.
  In such a way, the off-shell effects in the incoherent channel are
  evaluated consistently.

\begin{figure}[hb!]
     \parbox{.45\textwidth}{
    \includegraphics[width=0.45\columnwidth]{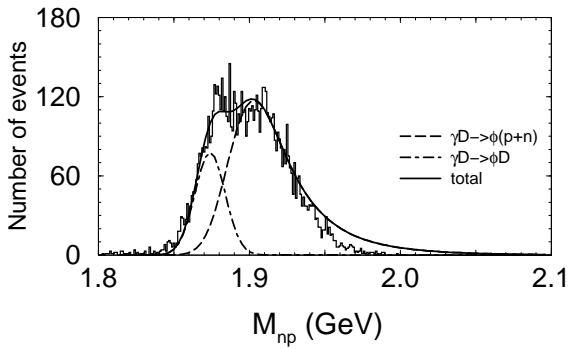}}
    \hfill
  \parbox{0.45\textwidth}{
   \caption{\small{Distribution of $[\gamma D, \phi]$ missing
   mass for the $\gamma D\to \phi X$ reaction. The histogram
   corresponds to
   the experimental data~\protect\cite{WenChen07add}.
   The theoretical curves are scaled by the factor
   3.7~$[\mu{\rm b/GeV}^3]^{-1}$ (see details in text).
   \label{Fig:9} }}}
   \end{figure}

 The differential cross sections of the incoherent $\phi$ meson
 photoproduction are displayed in Figs.~\ref{Fig:8}
 and~\ref{Fig:9}. Let us first discuss the differential missing mass
 distribution in the $\gamma D\to \phi X$ reaction ($X=D, np$) as a
 function of the $[\gamma D,\phi]$ missing mass and momentum transfer
 $t$.
 For the coherent and incoherent parts we use
 the common momentum transfer $t=t_D$. This means that the incoherent
 part must be multiply by the Jacobian
 $dt_X/dt_D =\sqrt{\lambda(s,M_X^2,M_\phi^2)/
 \lambda(s,M_D^2,M_\phi^2)}$. With regards to a comparison
 of our prediction to the experimental data,
 the experimental resolution must be included.
 Also, the cross section of the incoherent
 photoproduction is slightly modified.
 Therefore, we compare data with the missing
 mass distribution folded with a Gaussian distribution function
\begin{eqnarray}\label{RES1}
 \frac{d\sigma}{dM_X\,dt}&=&\int
 \frac{d\sigma}{dM\,dt}\, f(M_X-M)dM~,\nonumber\\
  f(M_X-M)&=&\frac{1}{\sigma\sqrt{2\pi}}
  \exp\left[-{\frac{(M_X-M)^2}{2\sigma^2}}\right]
\end{eqnarray}
with $\sigma=10$~MeV~\cite{WenChen07}, which imitates a finite
experimental resolution.

  In Fig.~\ref{Fig:8} we show the differential $[\gamma D,
  \phi]$ missing mass distribution in $\gamma D\to \phi X$
  reactions for different
  photon energies for forward photoproduction angle
  $\theta=0$.
  The position of the maximum of the incoherent part is marked
  by an arrow.
  One can see a strong energy dependence
  of (i) the absolute value of the cross section, (ii) the relative
  contributions of the coherent and incoherent processes,
  (iii) the position of the maximum of the incoherent part.
  At relatively large photon energies ($E_\gamma\sim
  2.5$~GeV) our model predicts a strong overlap of coherent and
  incoherent parts, and the coherent photoproduction amounts more than
  30\% of the total cross section. Our model seems to be an effective tool
  to isolate the coherent and incoherent parts with subsequent
  extraction of the $\phi$ photoproduction off the neutron.

  Fig.~\ref{Fig:9} exhibits the invariant mass distribution
  averaged within the interval $E_\gamma=1.5-2.4$~GeV
  together with experimental
  data~\cite{WenChen07add} given in units of events. The
  theoretical curves are scaled by the factor
  $3.7~[\mu{\rm b/GeV}^3]^{-1}$. The comparison is rather
  qualitative because we did not use the detailed acceptance corrections
  which may somehow modify the shape of the distributions.
  Nevertheless, the qualitative agreement between prediction and data
  seems to be quite encouraging.

\section{Spin density matrix elements}

 In this section we consider
 several important matrix elements of
 spin-density matrices $\rho^i_{\lambda\lambda'}$ ($i=0,1,2$)
 which determine the $\phi$ meson decay
 distribution in its rest frame in case of both
 unpolarized and linearly polarized
 photon beams. The  spin-density matrices are defined by
\begin{eqnarray}
 \rho^{0}_{\lambda\lambda'}&=&
 \frac{1}{N}\sum_{\alpha,\lambda_\gamma}
 T_{\alpha;\lambda,\lambda_\gamma}\,T^\dagger_{\alpha;\lambda',\lambda_\gamma}~,
 \nonumber\\
 \rho^{1}_{\lambda\lambda'}&=&
 \frac{1}{N}\sum_{\alpha,\lambda_\gamma}
 T_{\alpha;\lambda,-\lambda_\gamma}\,T^\dagger_{\alpha;\lambda',\lambda_\gamma}~,
 \nonumber\\
 \rho^{2}_{\lambda\lambda'}&=&
 \frac{i}{N}\sum_{\alpha,\lambda_\gamma}
 T_{\alpha;\lambda,-\lambda_\gamma}\,T^\dagger_{\alpha;\lambda',\lambda_\gamma}~.
  \label{Rho-1}
\end{eqnarray}
 The symbol $\alpha$ includes the polarizations of the
 incoming and  outgoing baryons, and the normalization factor
 has the standard form
 \begin{eqnarray}
 {N}=\sum_{\alpha,\lambda,\lambda_\gamma}
 T_{\alpha;\lambda,\lambda_\gamma}\,T^\dagger_{\alpha;\lambda,\lambda_\gamma}~,
 \label{Rho-2}
 \end{eqnarray}
 where $T_{\alpha;\lambda,\lambda_\gamma}$ is the total
 $\phi$ meson photoproduction amplitude.

 We perform  our consideration  in the $\phi$-meson rest
  frame with the quantization axis along the beam
  momentum, i.e. Gottfried-Jackson (GJ) system. Other possible
  choices
  are the helicity (H) system with quantization axis opposite to the
  recoil nucleon (deuteron) momentum in $\gamma p$ ($\gamma D$)
  reaction,
  and the Adair (A) system, where the quantization axis is along the beam
  direction in c.m.s.~\cite{Schilling70}.
  The GJ  system has some advantage because only here some of
  spin-density matrix elements have a clear physical meaning,
  e.g. as a measure of the helicity conserving processes
  or as an asymmetry between processes with natural and un-natural
  parity exchange in $t$ channel.

 Consider first
 the matrix element $\rho^0_{00}$.
 This matrix element determines the polar angular
 distribution of $\phi\to K\bar K$ decay
 \begin{eqnarray}
 W(\cos\Theta)= \frac{3}{2}
 \left(\rho^0_{00} +\frac12(1-3\rho^0_{00})
 \sin^2\Theta
 \right)~.
 \label{Rho-0}
 \end{eqnarray}
 In GJ system, $\rho^0_{00}$
 is the measure of the spin flip transition with
 $\lambda_\gamma=\pm1\to\lambda_\phi=0$.
 Thus, in case of a pure helicity conserving amplitude, which may be
 expressed as
\begin{eqnarray}
 T_{\alpha;\lambda_\phi,\lambda_\gamma}\simeq
 ({\bm\varepsilon}_{\lambda_\gamma}\cdot
 {\bm\varepsilon}^*_{\lambda_\phi})\,
 T^0_\alpha~,
 \label{Rho-3}
 \end{eqnarray}
  the photon polarization vector
  ${\bm\varepsilon}_{\lambda_\gamma}$ is transversal with respect to the ${\bm
  z}$ axis, and therefore spin-flip transitions
  $\lambda_\gamma=\pm1\to \lambda_\phi=0$ are forbidden and
  $\rho^0_{00}=0$, independently of the momentum transfer.
  In the helicity system, the photon polarization vector
  has a finite ${\bm z}$ component
\begin{eqnarray}
   {\varepsilon}_{\lambda_\gamma}{}_z=\frac{\lambda_\gamma}{\sqrt{2}}\sin\beta~,
   \label{Rho-4}
\end{eqnarray}
where $\beta$ is the angle between H and GJ systems
\begin{eqnarray}
\beta=\frac{v_\phi -\cos\theta}{v_\phi\cos\theta - 1}~,
  \label{Rho-5}
\end{eqnarray}
 and $v_\phi$ and $\theta$ are the $\phi$-meson velocity
 and the $\phi$ photoproduction angle in c.m.s., respectively.
 Foe relatively
 large momentum transfer, when $\sin\beta\simeq 1$, one gets
 a large value of $\rho^{0{\rm H}}_{00}$
\begin{eqnarray}
\rho^{0{\rm H}}_{00}\simeq \sin^2\beta~,
  \label{Rho-6}
\end{eqnarray}
 even for the helicity conserving amplitude. Conversely,
 one can imagine an amplitude which generates $\rho^{0{\rm GJ}}_{00}\simeq 1$
 (for example, take only the second term in Eq.~(\ref{P3})), and
 then  $\rho^{0{\rm H}}_{00}\simeq \cos^\beta\simeq0$.
In general, the spin-density matrices in H and GJ system are
related to each other as
\begin{eqnarray}
 \rho^{i\,{\rm H}}_{\lambda\lambda'}=\sum\limits_{\mu\nu}
 d^1_{\lambda\mu}(-\beta)\rho^{i\,{\rm GJ}}_{\mu\nu}
 d^1_{\nu\lambda'}(\beta)~.
  \label{Rho-7}
\end{eqnarray}

\begin{figure}[hb!]
    \includegraphics[width=0.4\columnwidth]{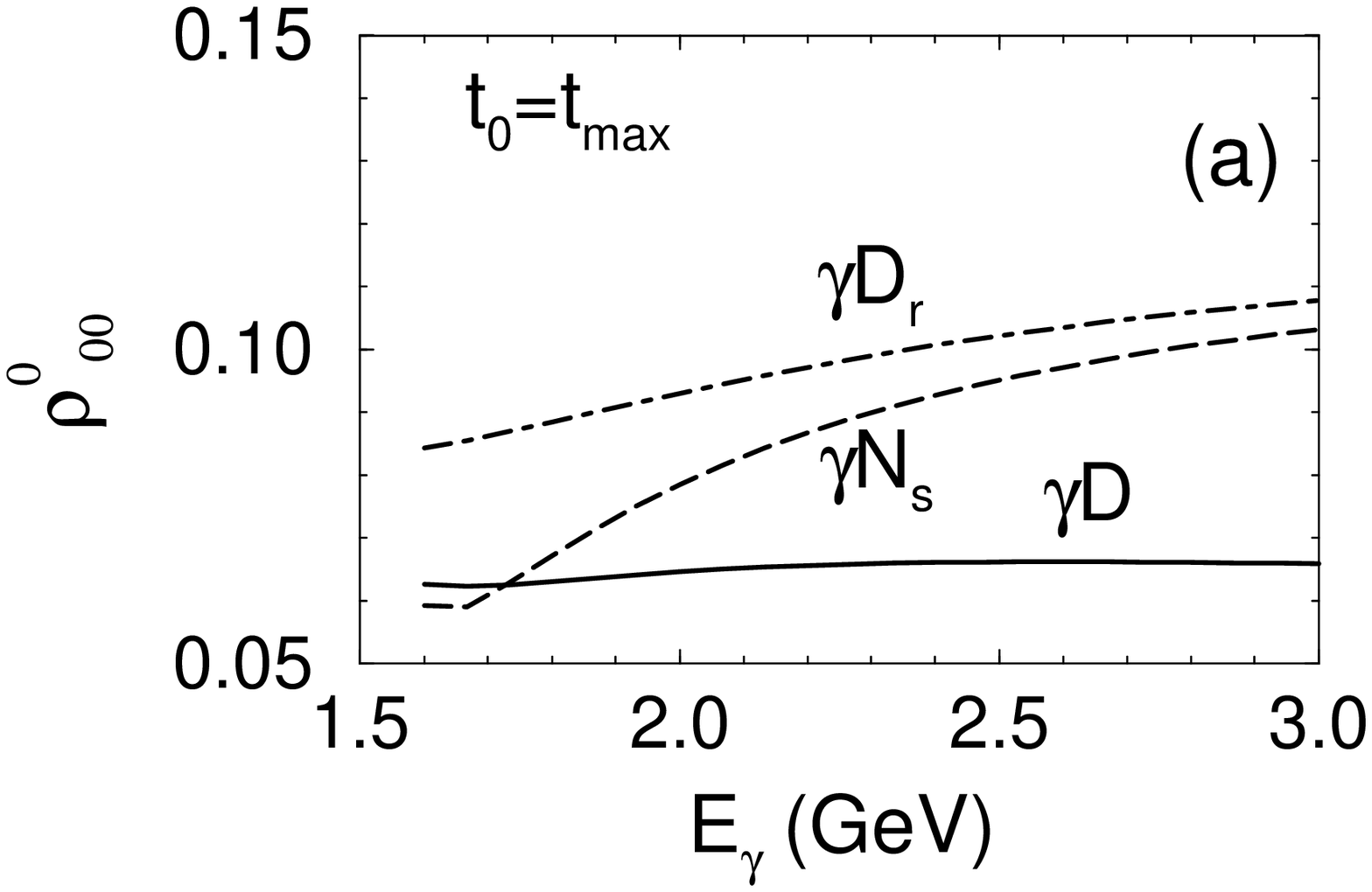}\qquad
    \includegraphics[width=0.4\columnwidth]{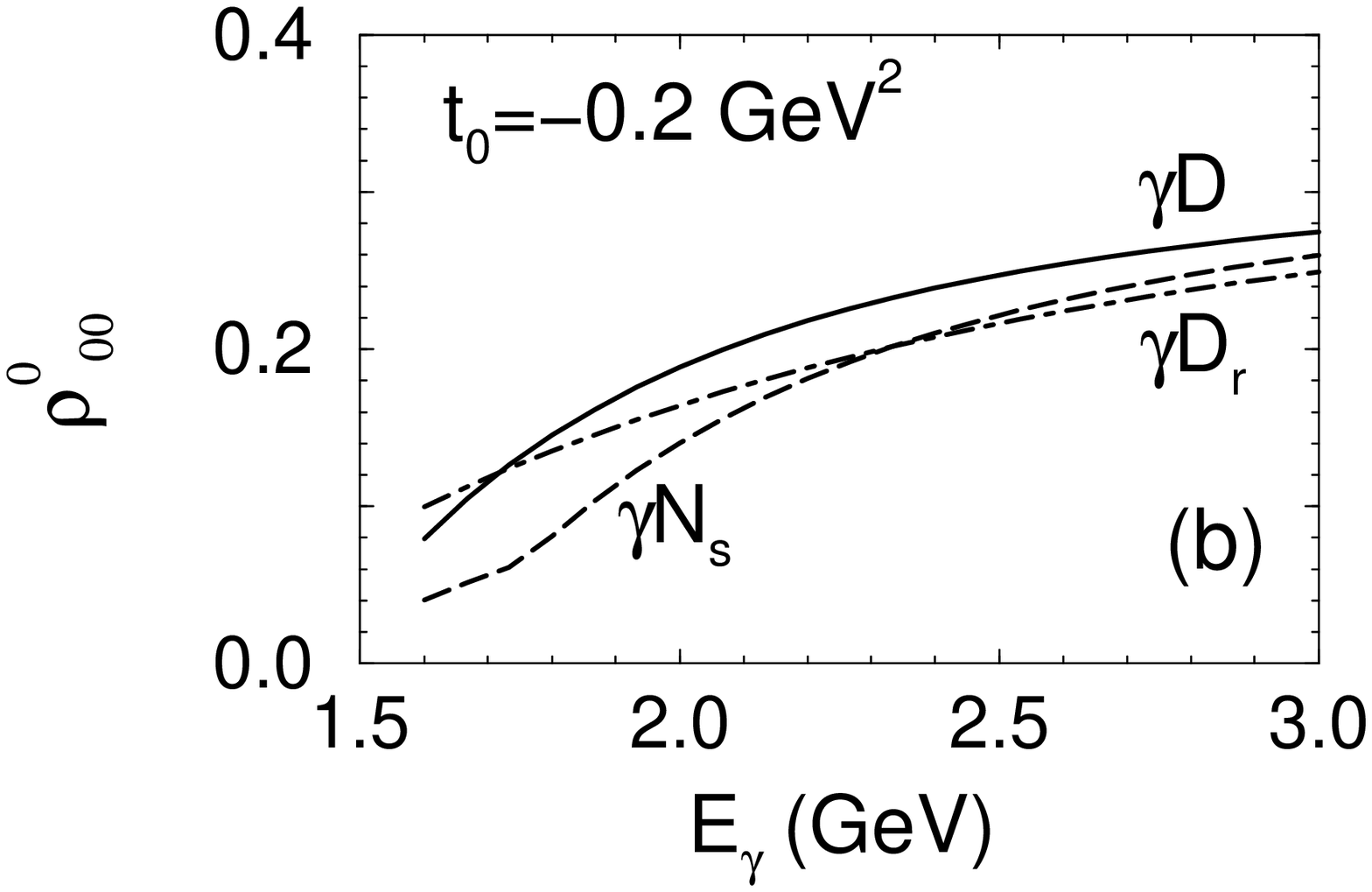}
   \caption{\small{The energy dependence of
    $\rho^0_{00}$ for the $\gamma D\to \phi D$ reaction.
    (a) and (b) correspond to $t_0=t_{\rm max}$ and
    $t_0=-0.2$~GeV$^2$, respectively, $\Delta_t=0.2$~GeV$^2$.
    The solid, dot-dashed and dashed curves correspond to
    the case of the explicit   $\gamma D$ reactions, the $\gamma D$
    reaction with reduced cross sections, and photoproduction off
    the free isoscalar nucleon, respectively
     \label{Fig:10}}}
   \end{figure}

 Let us first discuss the energy
 dependence of the spin-density matrix element
 $\rho^0_{00}$ in the GJ system for $\gamma p$, $\gamma n$ and $\gamma D$
 reactions.
 Following the experimental
 data, we calculate averaged $\rho$ matrices in
 the interval $|t|-|t_0|<\Delta_t$. The averaged $\rho$ matrices
 are defined as  ratios of averaged numerators and
 denominators ($N$) in Eqs.~(\ref{Rho-1}). In such a case, a direct
 comparison of the $\rho$ matrices for the coherent $\gamma D$ and
 for the $\gamma p$ reactions is hampered by the deuteron form
 factor. The deuteron form factor drops rapidly with increasing
 values $-t$ (see Fig.~\ref{Fig:4})
 and, therefore, the dominant contributions in $\gamma D$ and $\gamma p$
 reaction at the same values of $t_0$ and $\Delta_t$ come from  different
 momentum transfers $|\bar t_D|<|\bar t_p|$. This effect is particularly
 important for small values of $t_0\simeq t_{\rm max}$, where the slope of
 the deuteron form factor is rather steep. Thus, at relatively large
 energies, say  $E_\gamma\geq2$~GeV, the main contribution comes from
 $|\bar t_D|\simeq |t_{\rm 0}|\sim 0$, making the averaged $\rho$ matrices
 for the $\gamma D$ reaction practically constant.
 One can remove the effect of the deuteron form factor by scaling the
 product $TT^\dag$ in Eqs.~(\ref{Rho-1}) (or in the cross sections
 of the $\gamma D\to K^+K^-D$ reactions) by an inverse structure
 factor $Z(t)$ given by Eq.~(\ref{A8-1}). Such reduced $\rho$ matrices would be much
 closer to the $\rho$ matrices for the photoproduction off the "free"
 isoscalar nucleon. Fig.~\ref{Fig:10}
 illustrates effect of the deuteron form factor for the case of the
 $\gamma D$ reactions and the $\gamma D$
 reaction with reduced cross sections. The latter one is denoted as
 $\gamma D_r$. For completeness, we also
 show results for the $\phi$ photoproduction off the free isoscalar
 nucleon.
 One can see a large difference between predictions for
 $\gamma D$ reaction and the photoproduction off the
 free isoscalar nucleon at $t_0=t_{\rm max}$. In the first case, $\rho^0_{00}$
 is almost constant, whereas in the second case it increases with
 energy in the given energy interval. Such an increase for the $\gamma N$ reaction
 can be
 understood as follows. The finite value of  $\rho^0_{00}$
 is generated by the Pomeron exchange amplitude and
 is determined by the second (main) and third terms in Eq.~(\ref{P3}),
 whereas the total cross section
 is dominated by the first term.
 Neglecting spin conserving pseudoscalar meson exchange one can
 get the following analytical estimate of $\rho^0_{00}$ for GJ frame
  for the pure Pomeron exchange channel
 \begin{eqnarray}
 {\rho^{0}_{00}}_{\rm approx}\simeq
 \frac{2(2p_x^2 - t)k_\gamma^2}{(s-M_N^2)(s-M_N^2-M_\phi^2-t)}~,
  \label{Rho-7}
\end{eqnarray}
 where $p_x$ is the $x$ component of the nucleon momentum ($p_x=p_x'$),
 $k_\gamma$ is the photon energy, and $s$
 is the total energy squared in the $\gamma N$
 vertex. At fixed $t$, dependence on form factors
 in numerator and denominator for the $\gamma N$ reaction is canceled.
  The increase of $\rho^{0}_{00}$
 with energy, within the considered energy interval, is explained by
 a faster increase of the numerator (because of factor $p_x^2$)
 compared to the denominator at fixed $t$. At larger energies and small $|t|$
 this ratio and the corresponding matrix element decrease.
\begin{figure}[hb!]
\parbox{0.45\textwidth}{
    \includegraphics[width=0.45\columnwidth]{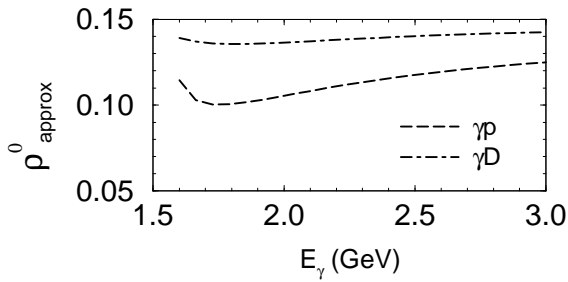}}
 \hfill
\parbox{0.4\textwidth}{
   \caption{\small{Estimates of $\rho^0_{00}$
   given by  Eq.~(\protect\ref{Rho-7}) for $\gamma p$
   and $\gamma D$ reactions.  \label{Fig:11}}}}
   \end{figure}

 The difference between reduced $\rho^0_{00}$ matrix element and the case of
 photoproduction off the isoscalar nucleon is explained by the
 difference in  $p_x$, $k_\gamma$, $t_{\rm max}$, and $s$
 for  $\gamma p$ and $\gamma D$ reactions.
 Actually, the kinematical
 variables in $\gamma N$ vertices in $\gamma p$ and  $\gamma D$
 reactions at fixed
 $E_\gamma$ and $t$ ($|t^d_{\rm max}|<|t^p_{\rm max}|$)
 are different and this
 difference is reflected in spin-density matrix elements.
 As an illustration,
 in Fig.~\ref{Fig:11} we exhibit results for $\rho^0_{00}$
 given as a ratio of the averaged numerator and denominator in
 Eq.~(\ref{Rho-7}) calculated
 for $\gamma p$ and $\gamma D$ kinematics.
 One can see some difference between the two cases
 caused by pure kinematics.

\begin{figure}[hb!]
    \includegraphics[width=0.4\columnwidth]{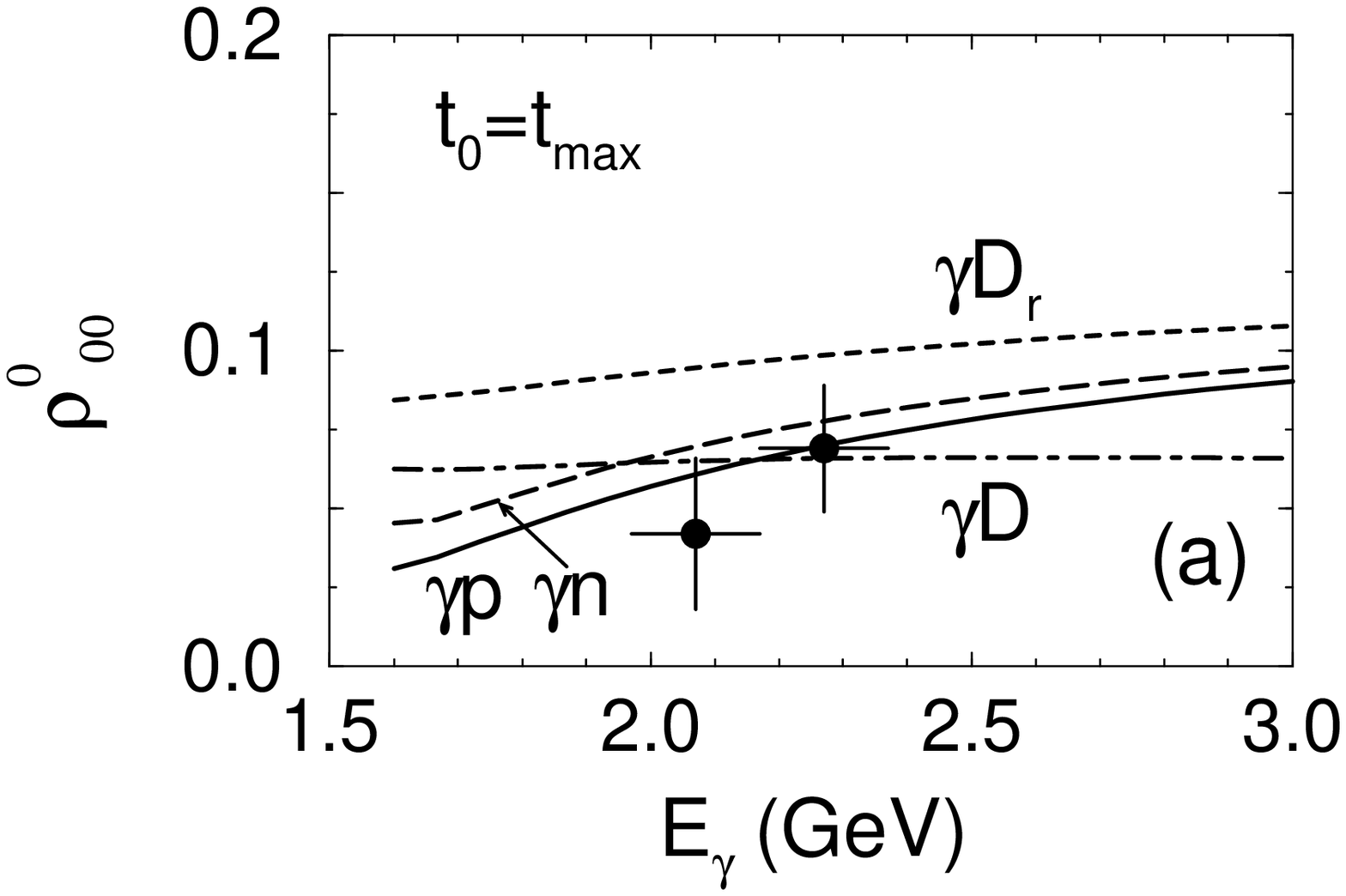}\qquad
    \includegraphics[width=0.4\columnwidth]{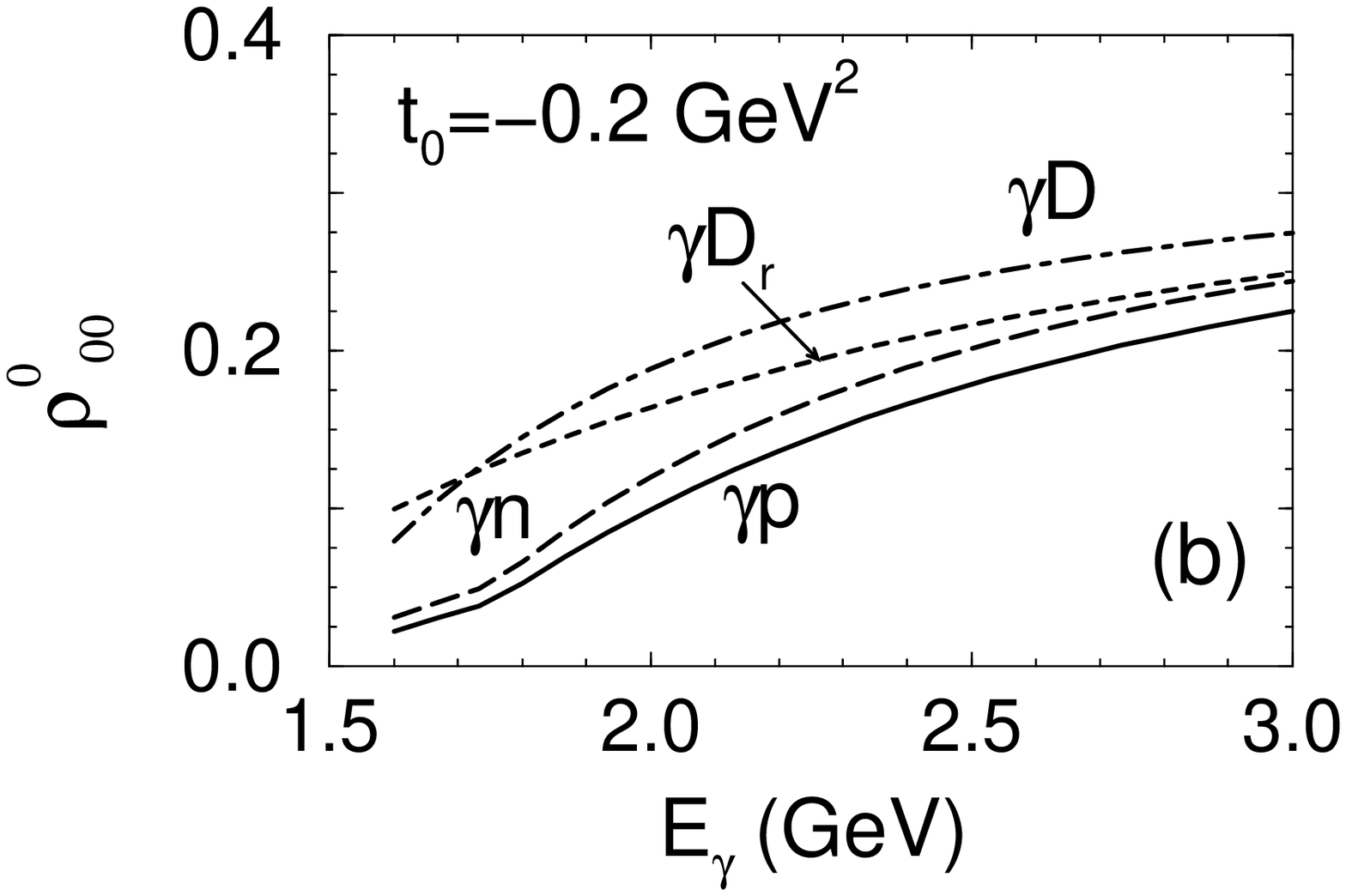}
   \caption{\small{The energy dependence of
    $\rho^0_{00}$.  (a) and (b)
    correspond to $t_0=t_{\rm max}$ and
    $t_0=-0.2$~GeV$^2$, respectively.
   The experimental data are taken from~\cite{Mibe05}.
     \label{Fig:12}}}
   \end{figure}

 The comparison of $\rho^0_{00}$ for the $\gamma p$, $\gamma n$ and
 $\gamma D$ reactions without and with scaling
 by $Z^{-1}(t)$ is shown in Fig.~\ref{Fig:12}.
 In Fig.~\ref{Fig:12}~(a) we show the result for
 forward photoproduction angles with
 $t_0=t_{\rm max}$ ($\theta=0$),
 together with available experimental data~\cite{Mibe05}.
 In Fig.~\ref{Fig:12}~(b)
 we choose the case of a larger momentum transfer
 with $t_{0}=-0.2$~GeV$^2$ for each energy.
  One can see a monotonic increase of $\rho^0_{00}$ with energy, and
  the inequality $\rho^0_{00}(\gamma p)< \rho^0_{00}(\gamma n)<\rho^0_{00}(\gamma
  D_r)$ holds.
 Some enhancement of $\rho^0_{00}$ in  $\gamma n$ reactions is
 explained by the destructive interference in the $\pi - \eta$ meson exchange
 amplitude which leads to a decrease of the helicity conserving
 terms in the full amplitude. Therefore, the relative contribution of
 the spin-flip terms in the $\gamma n$ reaction (cf. Eq.~(\ref{P3}))
 would be larger. In case of the $\gamma D_r$ reaction, together with
 a total suppression of $\pi$ meson exchange, $\rho^0_{00}$
 increases  additionally because of some difference in kinematics,
 as discussed above.

\begin{figure}[hb!]
\parbox{0.4\textwidth}{
    \includegraphics[width=0.35\columnwidth]{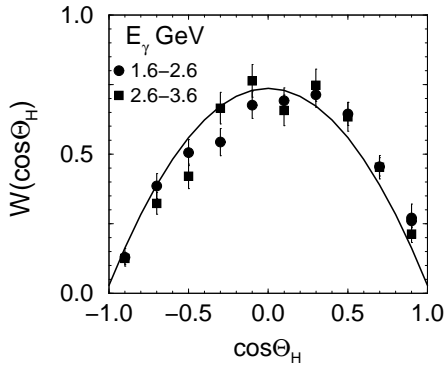}}
\hfill
    \parbox{0.5\textwidth}{
   \caption{\small{
   The angular distribution $W(\cos\Theta)$
   for the $\gamma D\to \phi D\to K^+K^- D$ reaction
   in the helicity frame at $E_\gamma=3.1$~GeV and
   $-t_0=0.3$~GeV$^2$.
   The experimental data for two energy intervals
   and $|t|=0.35-0.8$~GeV$^2$ are taken from~\protect\cite{Mibe07}.
    \label{Fig:13}}}}
   \end{figure}

\begin{figure}[hb!]
\parbox{0.4\textwidth}{
    \includegraphics[width=0.35\columnwidth]{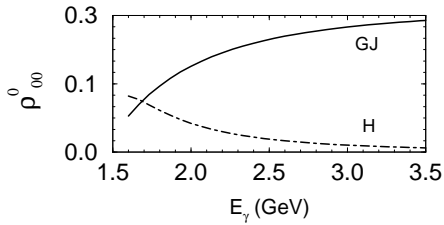}}
\hfill
    \parbox{0.5\textwidth}{
   \caption{\small{Spin-density matrix elements $\rho^0_{00}$
   in the helicity and Gottfried-Jackson frames
   at $|t|-|t_0|<0.2$~GeV$^2$ and $t_0=-0.2$~GeV$^2$.
     \label{Fig:14}}}}
   \end{figure}

 In Fig.~\ref{Fig:13} we exhibit the angular distribution $W(\cos\Theta)$
 in the  $\gamma D\to \phi D\to K^+K^- D$ reaction
 in the {\it helicity} frame for $E_\gamma=3.1$~GeV and for $t_0=-0.3$~GeV$^2$
 together with available experimental
 data~\cite{Mibe07} given in this frame. The shown experimental data
 are obtained in two energy bins with $E_\gamma=1.6-2.6$ and
 $2.6-3.6$~(GeV)and momentum transfer $|t|=0.35-0.8$~GeV$^2$.
 In our calculation the momentum transfer is in the range
 $|t|=0.3-0.5$~GeV$^2$, which corresponds to
 an upper bound of the momentum transfer acceptable
 for our model for the $\gamma D\to \phi D$ reaction
 with single scattering processes.
 Nevertheless, one can see a reasonable agreement between
 calculation and data.
 Note that this distribution is
 different in different frames because of the frame dependence of
 the $\rho$ matrices. As an example, in Fig.~\ref{Fig:14}
 we show the energy dependence of $\rho^0_{00}$ for
 the $\gamma D\to \phi D$ reaction in H and GJ
 frames at $|t|-|t_0|<0.2$~GeV$^2$ and $-t_0=0.2$~GeV$^2$.

\begin{figure}[hb!]
    \includegraphics[width=0.4\columnwidth]{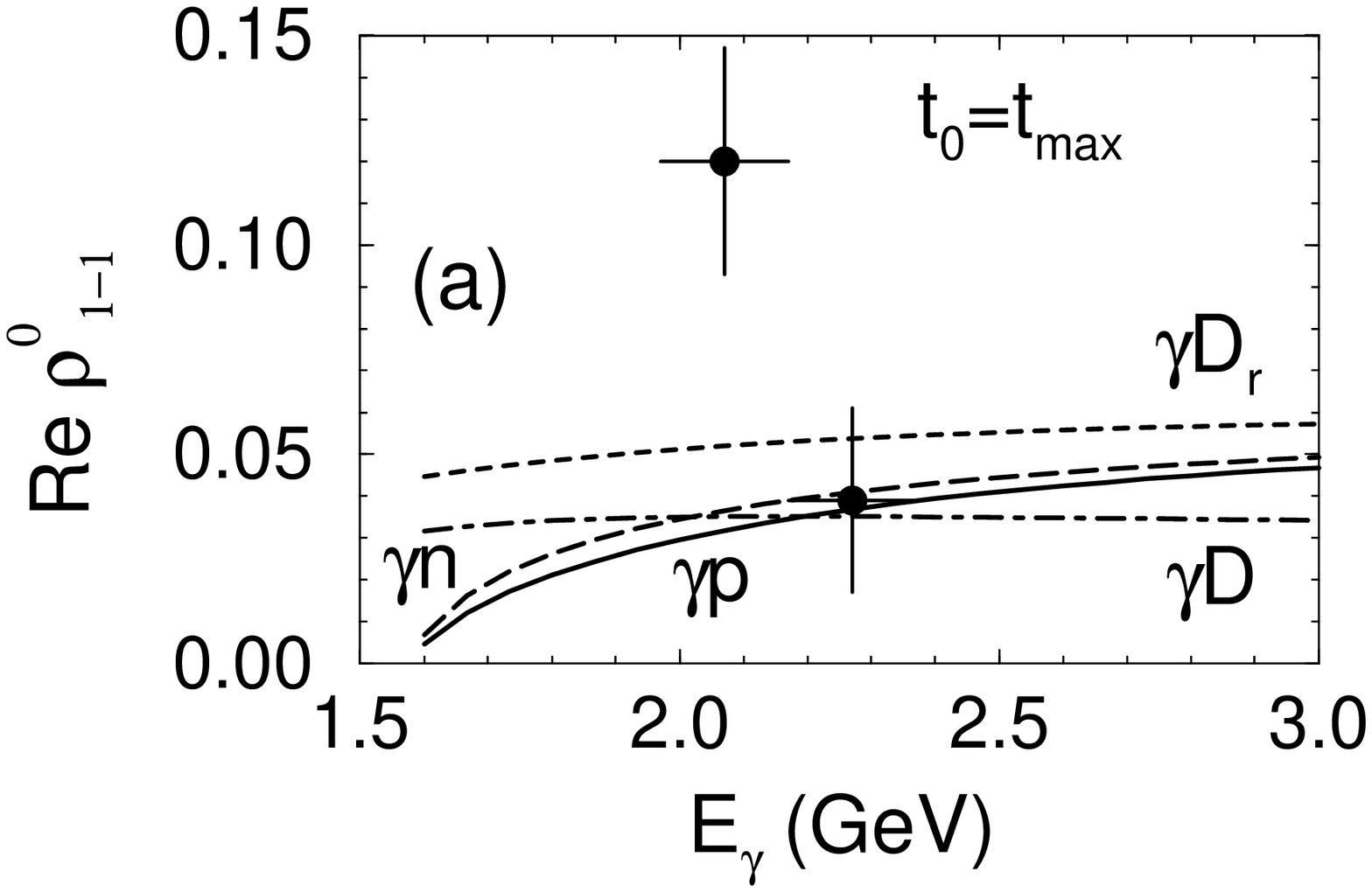}\qquad
    \includegraphics[width=0.4\columnwidth]{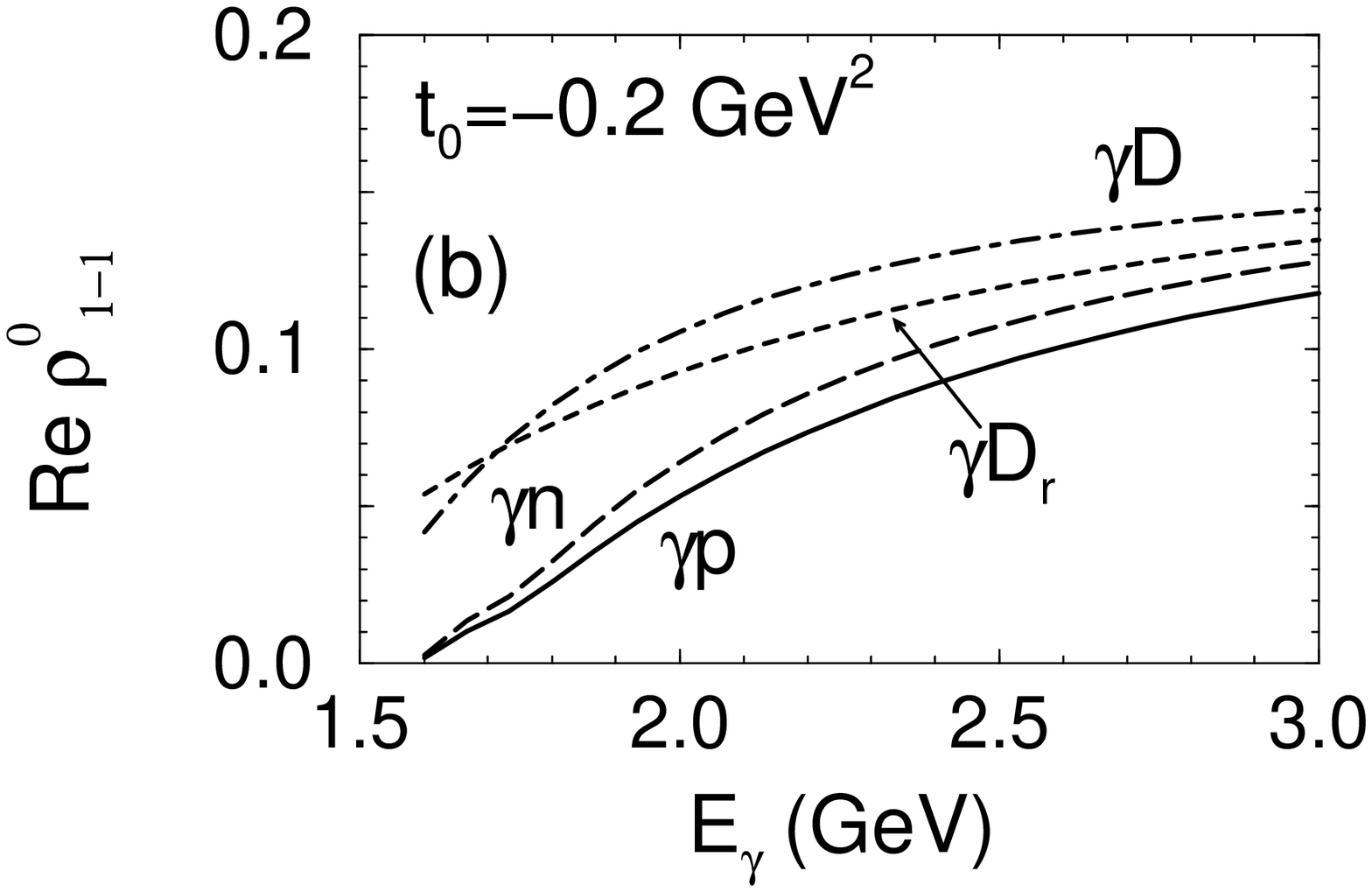}
   \caption{\small{
    The same as in Fig.~\protect\ref{Fig:12},
    but for
    ${\rm Re}\rho^0_{1-1}$.
     \label{Fig:15}}}
   \end{figure}

 The energy dependence of the spin-density matrix element
${\rm Re}\rho^0_{1-1}$ is displayed in
 Fig.~\ref{Fig:15}.  This matrix element determines the azimuthal
 angle distribution of $\phi\to K\bar K$ decay in reactions with
 an unpolarized photon beam
\begin{eqnarray}
 W^0(\Phi)=
 \frac{1}{2\pi}(1-2{\rm Re}\rho^0_{1-1}\cos2\Phi)~.
  \label{Rho-8}
\end{eqnarray}
 The matrix element $\rho^0_{1-1}$ is proportional to the relative
 contribution of processes with double spin
 transition where $\lambda_\gamma=\pm1\to\lambda_\phi=\mp1$. In our
 model, these transitions are generated by the last term in
 Eq.~(\ref{P3}). In  Fig.~\ref{Fig:15}~a (b) we show results for
 $|t|-|t_0|<0.2$~GeV$^2$ with $t_0=t_{\rm max}$ ($-0.2$~GeV$^2$),
 together with available experimental data~\cite{Mibe05}.
 The reason of the
 inequality $\rho^0_{1-1}(\gamma p)< \rho^0_{1-1}(\gamma n)<\rho^0_{1-1}(\gamma
  D)$ is similar to that in the previous case of single spin-flip
  transitions.

 The matrix elements $\rho^{1,2}_{1-1}$ are related to the
 asymmetry of transitions with natural (first term of
 Eq.~(\ref{P3})) and un-natural ($\pi,\eta$) parity exchange.
 They determine the $\phi$ meson decay distribution in case
 of linearly polarized photons as a function of the angle
 between azimuthal decay angle ($\Phi$)
 and the angle of the polarization plane ($\Psi$)
\begin{eqnarray}
 W^L(\Phi-\Psi)=
 \frac{1}{2\pi}(1+2P_\gamma\bar\rho^1_{1-1}\cos2(\Phi-\Psi))~,
  \label{Rho-9}
\end{eqnarray}
where $P_\gamma$ is the strength of polarization and
 \begin{eqnarray}
 \bar\rho^1_{1-1} =
\frac12(\rho^1_{1-1} -{\rm Im}  \rho^2_{1-1})\simeq\rho^1_{1-1} ~.
  \label{Rho-10}
\end{eqnarray}

\begin{figure}[hb!]
    \includegraphics[width=0.4\columnwidth]{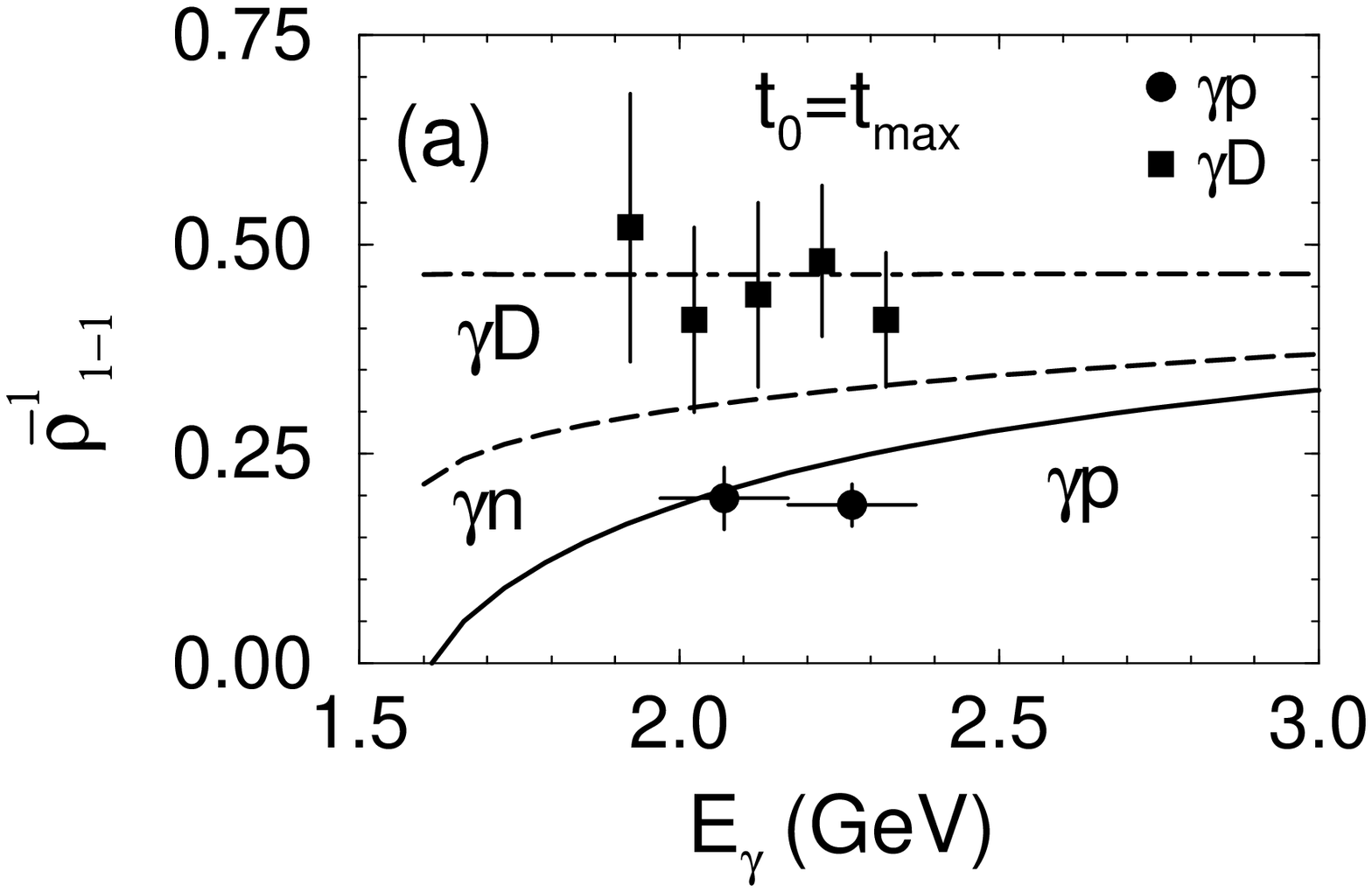}\qquad
    \includegraphics[width=0.4\columnwidth]{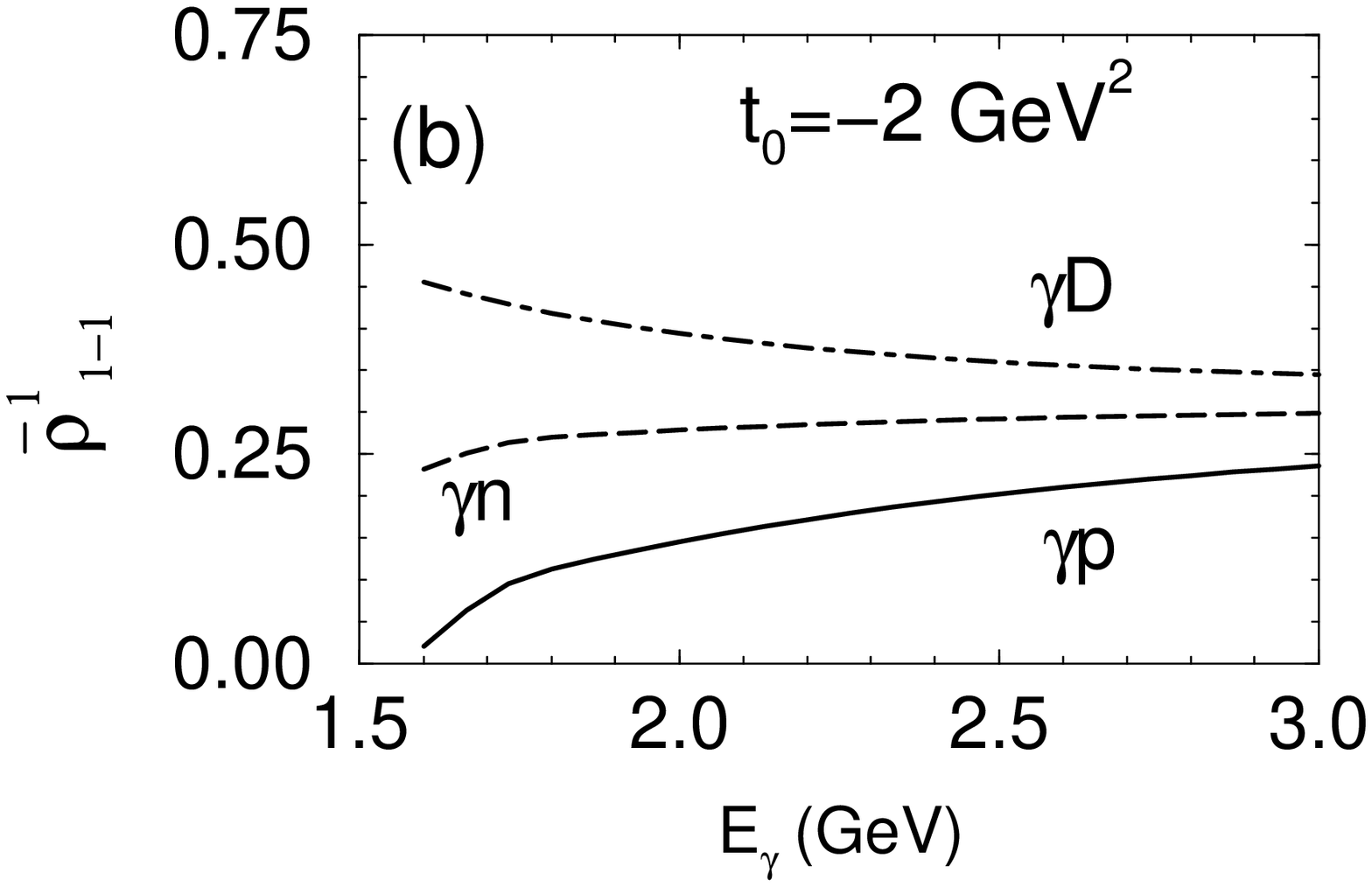}
   \caption{\small{The same as in Fig.~\protect\ref{Fig:12},
    but for
    $\rho^1_{1-1}$.
   The experimental data are taken from~\protect\cite{Mibe05,WenChen07}.
     \label{Fig:16}}}
   \end{figure}

 The energy dependence of the spin-density matrix element
 $\bar\rho^1_{1-1}$ is shown in Fig.~\ref{Fig:16}
 together with the experimental data~\cite{Mibe05,WenChen07}.
 In this case, the effect of the deuteron form factor is rather weak
 and we do not display results for the reduced matrix element.
 For pure natural (un-natural)
 parity exchange it is equal 0.5 (-0.5). Qualitatively, within experimental
 accuracy, the result of our calculation is consistent with the data.
 Sizeable deviations
 of $\bar\rho^1_{1-1}$ from 0.5 in the $\gamma p$
 reaction at low energy is explained by a large
 contribution of the
 $\pi,\eta$ exchange processes. Thus, at $E_\gamma\simeq2$~GeV
 they contribute on the level of 30\% to the total cross section.
 In $\gamma n$ and $\gamma D$ reactions the pseudoscalar exchange
 contributions  are suppressed, shifting $\bar\rho^1_{1-1}$ towards
 0.5.
\begin{figure}[hb!]
    \includegraphics[width=0.4\columnwidth]{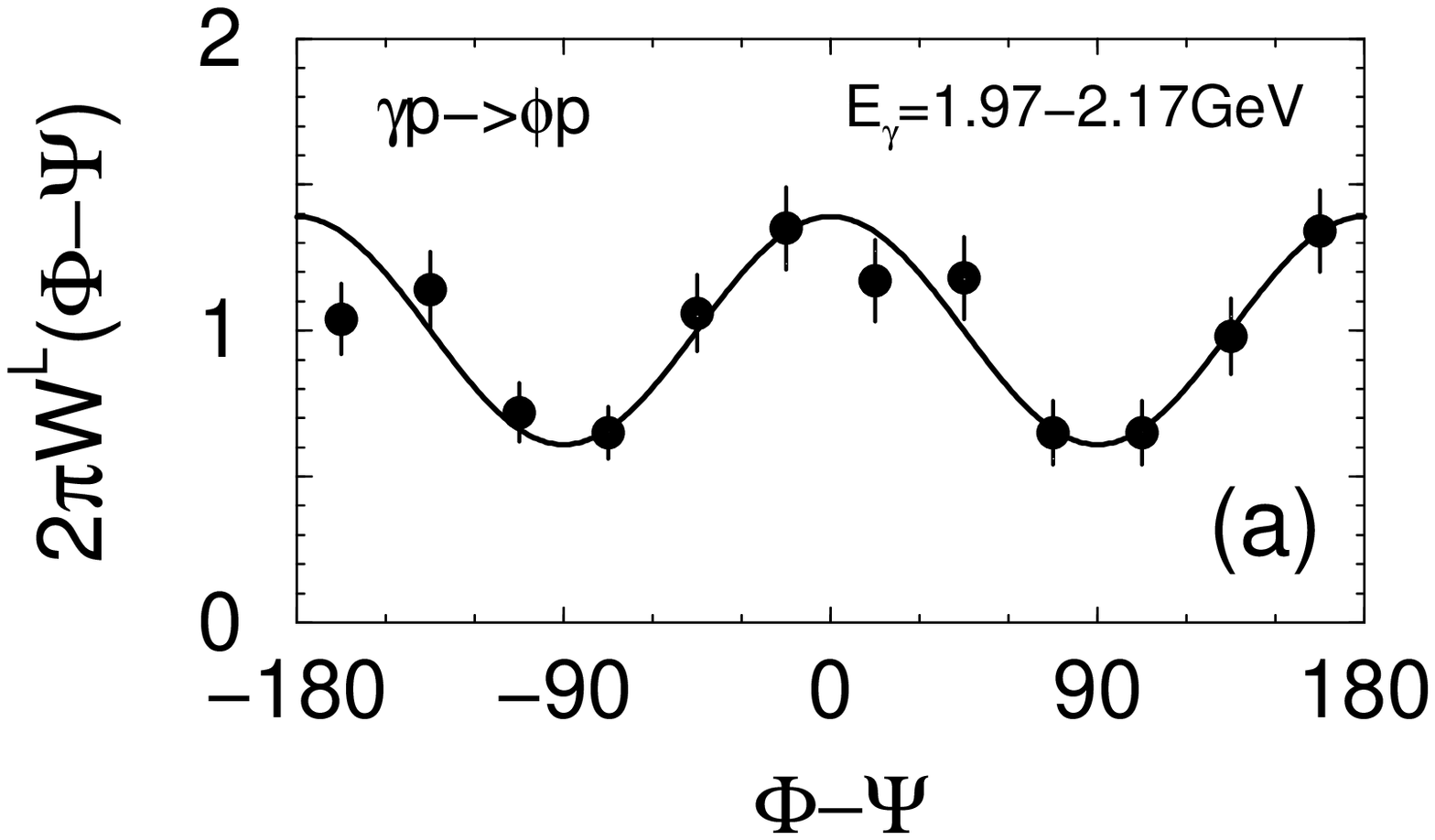}\qquad
    \includegraphics[width=0.4\columnwidth]{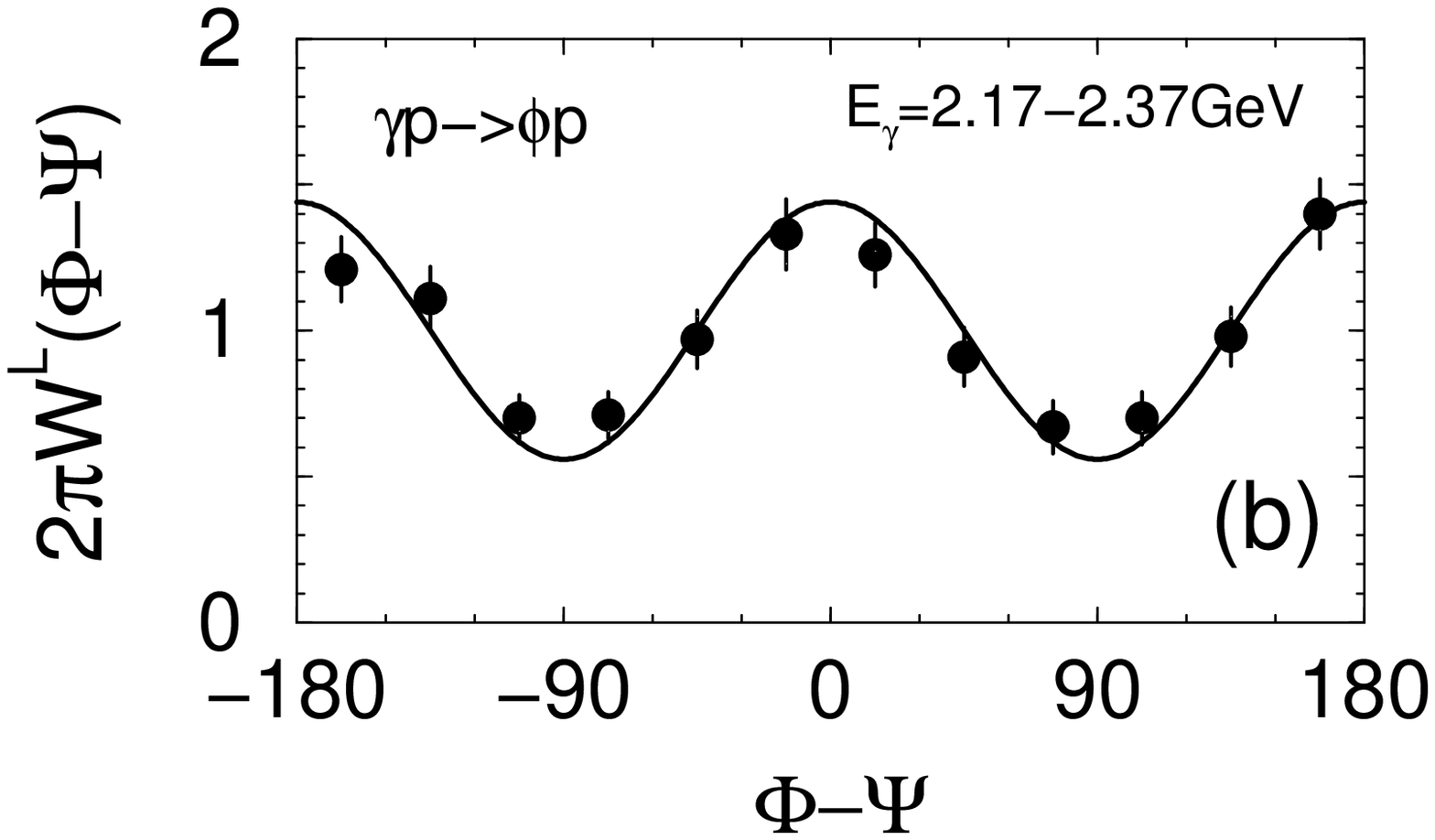}
   \caption{\small{
     The angular distribution  $W^L(\Phi-\Psi)$
     for the reaction $\gamma p\to \phi p\to p K^+K^-$
     at $|t|-|t_{\rm max}|\leq0.2$ GeV$^2$.
     (a) and (b) correspond to the
     energy intervals $E_\gamma=1.97-2.17$ and $2.17-2.37$ (GeV),
     respectively. The experimental data are from
     ~\protect\cite{Mibe05add}.
      \label{Fig:17}}}
   \end{figure}

 For completeness, we also present the angular distribution $W^L(\Phi-\Psi)$
 of Eq.~(\ref{Rho-9}) for different cases.  Figure~\ref{Fig:17}
 exhibits this angular distribution for the reaction
 $\gamma p\to \phi p\to p K^+K^-$
 at $|t|-|t_{\rm max}|\leq0.2$ GeV$^2$
 in two energy intervals $E_\gamma=1.97-2.17$ and $2.17-2.37$ (GeV)
 with beam polarization $P_\gamma=0.86$ and $0.90$, respectively,
 together with available experimental data~\cite{Mibe05add}.
 One can see a reasonable agreement between our calculation and the
 experiment.
\begin{figure}[hb!]
    \includegraphics[width=0.4\columnwidth]{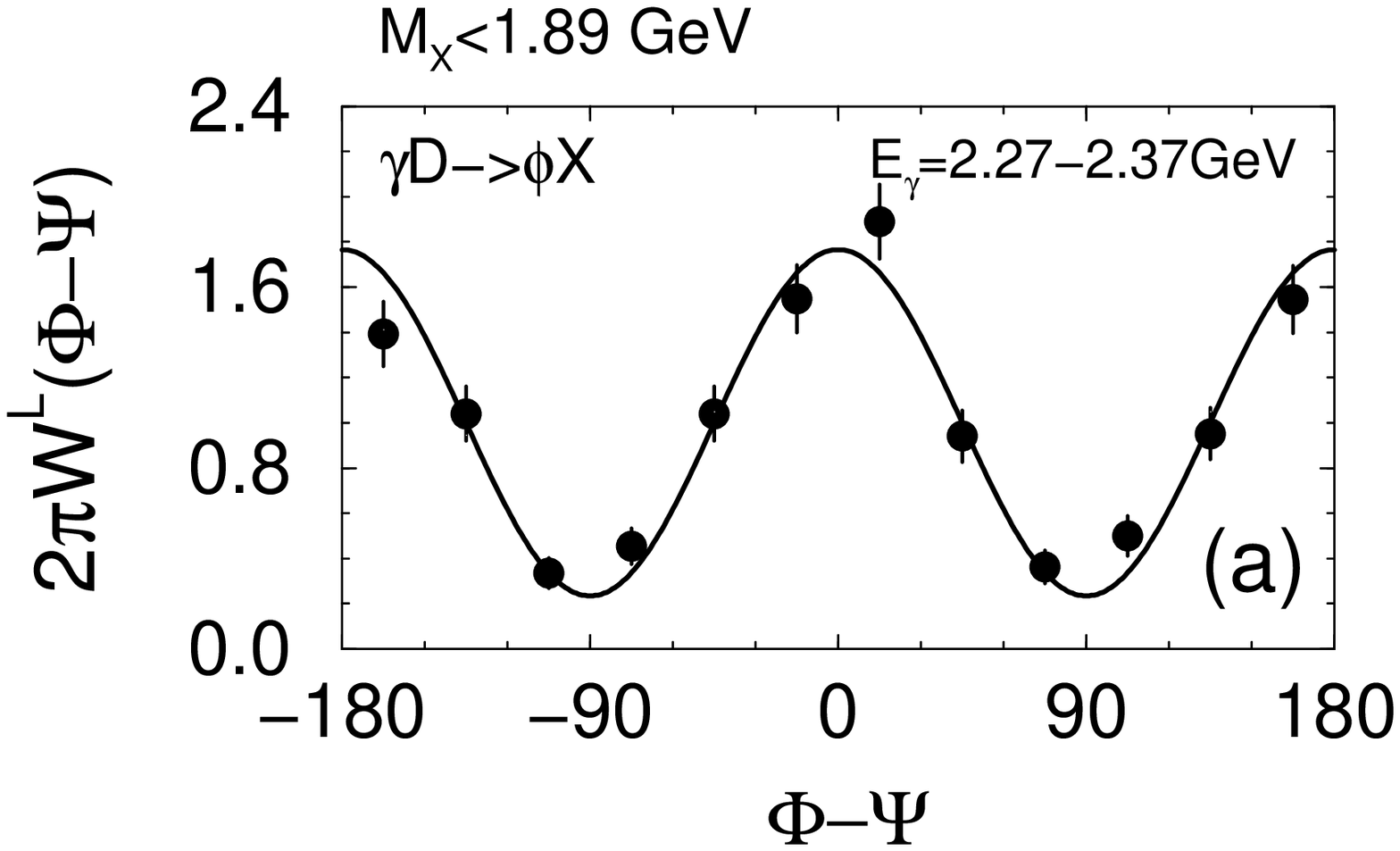}\qquad
    \includegraphics[width=0.4\columnwidth]{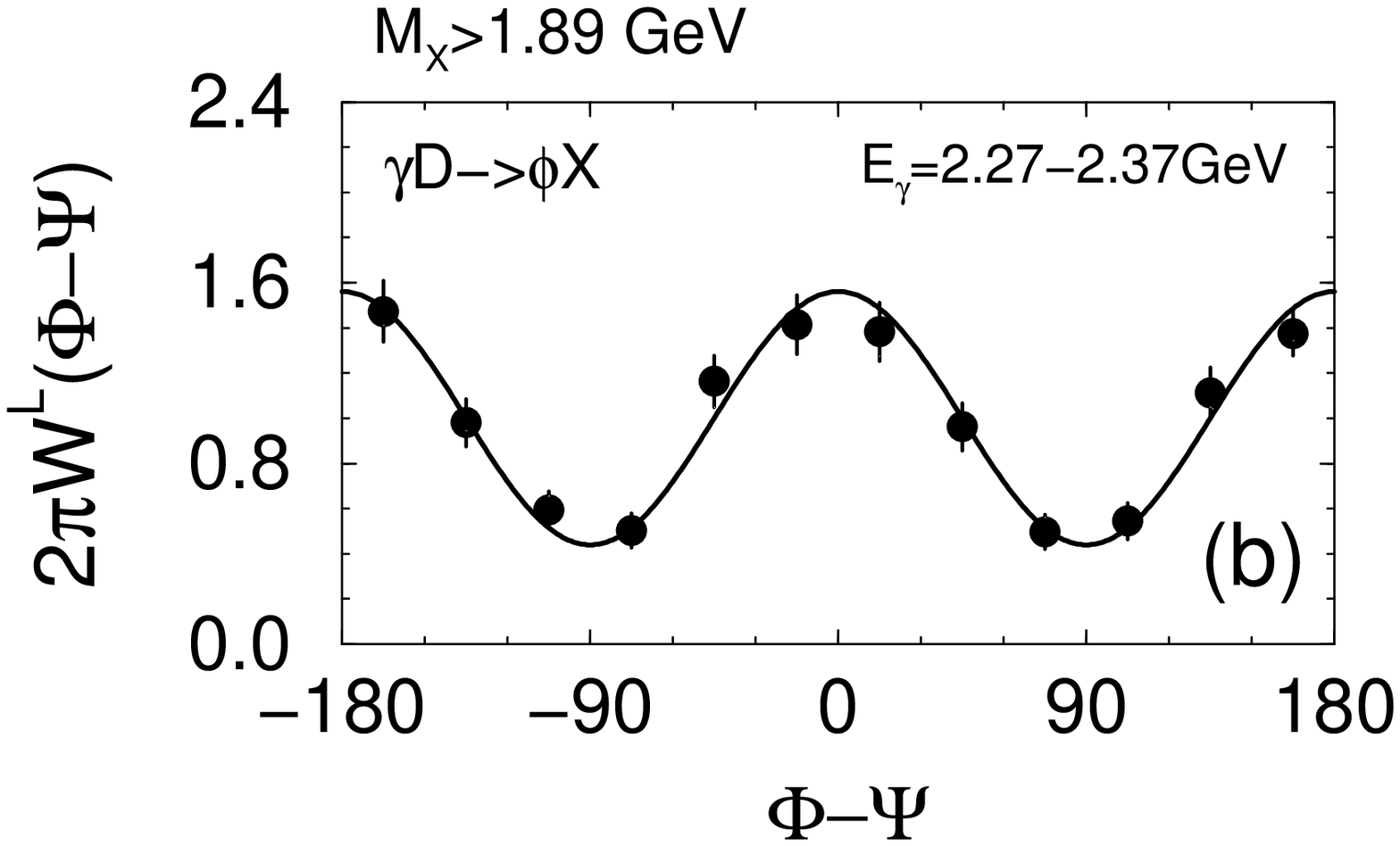}
   \caption{\small{
     The angular distribution  $W^L(\Phi-\Psi)$
     for the reaction $\gamma D\to \phi X\to X K^+K^-$
     ($X=D,np$)
     at $|t|-|t_{\rm max}|\leq0.1$ GeV$^2$.
     (a) and (b) correspond to
     the $[\gamma D,\phi]$ missing mass smaller or larger
     than 1.89 GeV,
     respectively. The experimental data are from
     ~\protect\cite{WenChen07}.
      \label{Fig:18}}}
   \end{figure}

 The angular distribution $W^L(\Phi-\Psi)$
 for the inclusive $\gamma D\to \phi X$ ($X=D,np$) reaction
 is displayed in Fig.~\ref{Fig:18} together with
 the experimental data of Ref.~\cite{WenChen07}.
 This distribution is calculated using the model, developed in Sec.~III.
 The left (a) and right (b) panels  correspond to events with $[\gamma D,\phi]$
 missing mass smaller or larger than $M_{\rm cut}=1.89$~GeV,
 respectively. In the first case the contributions come both from
 the coherent and incoherent $\phi$ meson photoproduction.
 The "effective" $\bar\rho^1_{1-1}$ matrix element is expressed as a sum
 \begin{eqnarray}
 \bar\rho^{1L}_{1-1}{}_{\rm eff} = {\bar\rho^1_{1-1}}{}_D\,P_{CH} +
 {\bar\rho^1_{1-1}}{}_{np}\,(1-P_{CH})~,
  \label{Rho-11}
\end{eqnarray}
where $P_D$ is the relative weight of the coherent channel, and
$\rho_{np}=(\rho_n+\rho_p)/2$ is the $\rho$ matrix for the
quasi-free nucleon. In the second case, the contribution of the
coherent channel is negligible and we get
 \begin{eqnarray}
{ \bar\rho^{1R}_{1-1}}{}_{\rm eff} \simeq
{\bar\rho^1_{1-1}}{}_{np}.
  \label{Rho-11}
\end{eqnarray}
In Fig.~\ref{Fig:18} we show result for $|t|-|t_{\rm
max}|<0.1$~GeV$^2$ and the energy bin with
$E_\gamma=2.27-2.37$~GeV~\cite{WenChen07}. Here, the beam
polarization is $P_\gamma=0.935$ and the model predicts
$P_{CH}\simeq0.67$. One can see a sufficient agreement between the
theoretical curves and the data. A similar agreement holds for the
other energy bins, too.

The agreement between the experimental data and the calculations
for the $K^+K^-$ angular distributions in $\gamma p$ and $\gamma
np$ reactions means that the model describes correctly the $\phi$
photoproduction off the neutron, and in particularly, supports our
choice of the pseudoscalar channel with a small contribution of
the $\eta$ meson exchange.

\begin{figure}[hb!]
    \includegraphics[width=0.4\columnwidth]{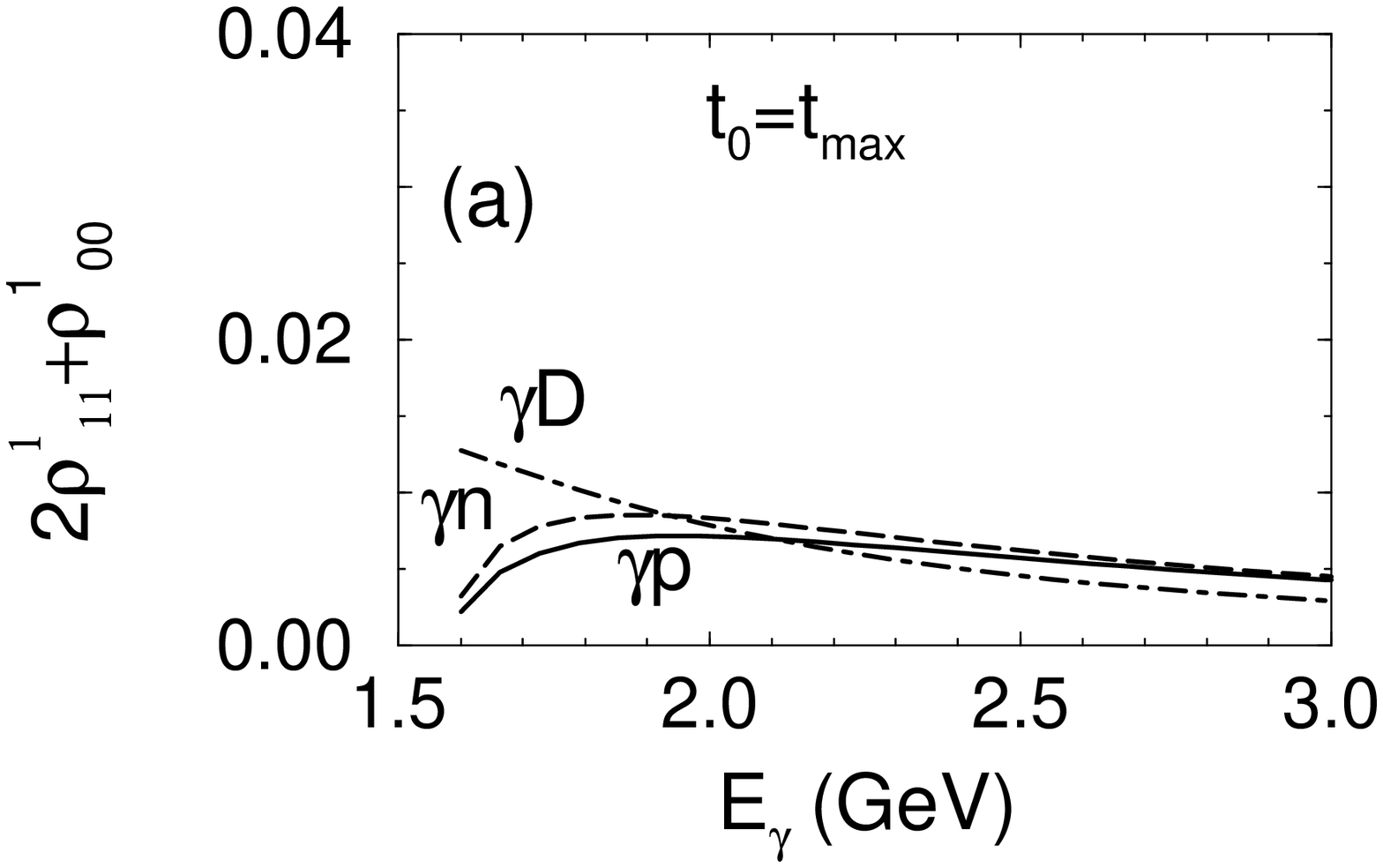}\qquad
    \includegraphics[width=0.4\columnwidth]{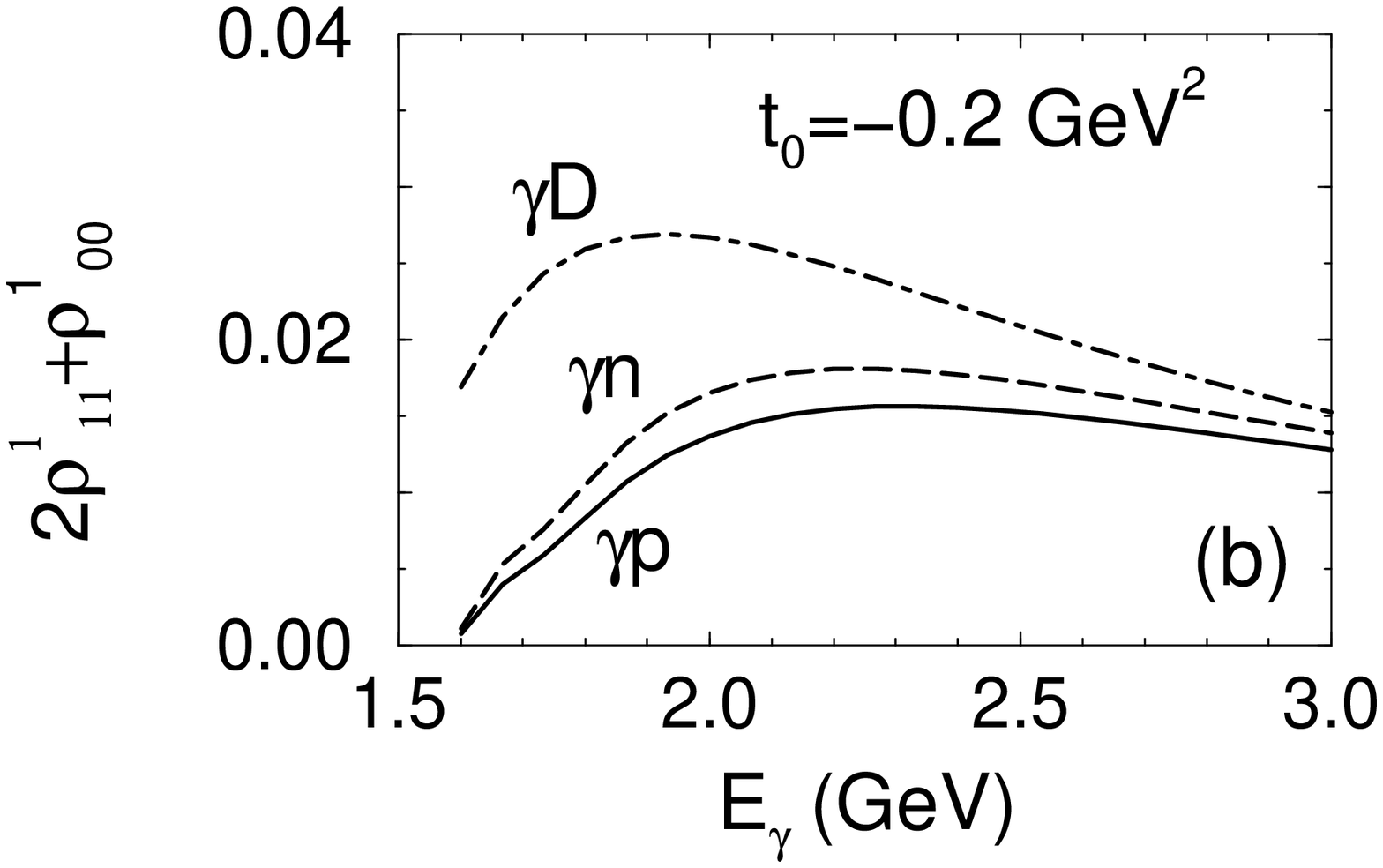}
   \caption{\small{
     The energy dependence of
    $2\rho^1_{11}+\rho^1_{00}$.  The left panel and the right
    panels correspond to $t_0=t_{\rm max}$ and
    $t_0=-0.2$~GeV$^2$, respectively.
      \label{Fig:19}}}
   \end{figure}
The sum $\rho^1_M\equiv 2\rho^1_{11} + \rho^1_{00}$ determines the
$\phi$ meson decay distribution as a function of the angle between
production and beam polarization planes
\begin{eqnarray}
 W^L(\Psi))=
 \frac{1}{2\pi}(1+2P_\gamma\rho^1_M\cos2\Psi)~.
  \label{Rho-11}
\end{eqnarray}
 It is important that parity conservation requires
 $\rho^1_{\mu\nu}=(-1)^{\mu-\nu}\rho^1_{-\mu-\nu}$~\cite{Schilling70},
 which makes $\rho^1_M$ invariant under rotation of the
 coordinate frame in the production plane. This means that
 $\rho^{1\rm GJ}_M=\rho^{1\rm H}_M=\rho^{1\rm A}_M$.
 Therefore, it is natural that this invariant function
 determines the distribution which depends only on the
 beam polarization.

Since $\rho^1_M$ is proportional to a combination of single and
double spin-flip transition amplitudes, its absolute value is
small.  The energy dependence of $\rho^1_M$ is shown in
Fig.~\ref{Fig:19}. One can see some increase of $\rho^1_M$ when
going from $t_0=t_{\max}$ (a) to $t_0=-0.2$~GeV$^2$ (b).  This is
explained by an increasing contribution of spin-flip transitions
with $|t|$.

\section{summary and discussion}

  We studied different aspects of coherent and
  incoherent $\phi$ meson photoproduction off the deuteron
  at forward photoproduction angles
  with the aim to check whether
  the recent experimental data require the
  inclusion of  some exotic channels discussed in literature.
  For this purpose we re-analyzed the elementary $\gamma p\to\phi p$
  reaction in order to use it as an input for our study.
  The corresponding amplitude in the diffractive region is expressed
  as a sum of Pomeron and pseudoscalar exchange channels.
  The first one represents a slightly modified Donnachie-Landshoff
  Pomeron exchange amplitude, whereas the second one is
  the coherent sum of the $\pi$ and $\eta$ meson exchange channels.
  In present work the contribution of the $\eta$ exchange channel
  is relatively weak, and correspondingly, the $\pi$ exchange is enhanced
  in order to get the proper relative contributions of the channels with
  natural and un-natural parity exchange.
  The Donnachie-Landshoff model is designed for high energy and it is not clear
  whether it can be applied at low energies, and close to the threshold
  as well.

  We performed a detailed analysis of the differential cross section
  of the $\gamma p\to\phi p$ reaction at $E_\gamma\sim2$~GeV and
  obtained a reasonable agreement between the model predictions and
  the available experimental data in diffraction region.
  At larger  momentum transfer our model underestimates recent
  data of LEPS and SAPHIR but is  quite reasonable for the Bonn and
  JLab data up to $t=0.8$~GeV$^2$.
  On the other hand, the Pomeron exchange model, motivated by the two-gluon dynamics,
 contains terms responsible for single and double spin-flip
 transitions. The model predictions for the spin-density matrix
 elements being sensitive to the spin-flip transitions are in agreement
 with available data for the $\gamma p$ reaction at $E_\gamma\sim2$~GeV,
 which also decreases the space left for possible exotic channels.
 Therefore we can conclude, that for a clear understanding a possible
 manifestation of an exotic channel one needs a complete set  of $t$
 dependences for unpolarized cross sections and polarization observables at different
 energies.

  We developed a model for the coherent and incoherent $\phi$
  meson photoproduction off the deuteron and performed again
  a detailed analysis of the existing data.
  The slope of the differential cross section of the coherent
  $\phi$ meson photoproduction is defined by the
  corresponding slope of the elementary $\gamma N$ reaction and by the deuteron
  form factor. We found a quite reasonable agreement between the
  model prediction and the experimental data in  the diffractive region
  and some underestimate at large $|t|\sim 0.4$~GeV$^2$, which favor
  the contributions of more complicated channels, for example, double scattering
  processes. But on the other hand, the model calculation of the $\phi\to K^+K^-$
  decay distribution, $W(\cos\Theta)$, at $|t|\simeq0.4$~GeV is in a good agreement
  with the experimental data, which, to some extent, support the
  single scattering model in this region of $t$. Therefore, the
  remaining
  difference between theory and experiment at $|t|\sim 0.4$~GeV$^2$
  requires further investigation.

  The model fairly well describes the energy dependence
  of  the cross section of the $\gamma D\to \phi
  D$ reaction at $\theta=0$ without any hint to a bump-like
  behavior.

  We performed detailed and combined investigation of several important
  spin density matrix elements for $\gamma p\to\phi p$,
  coherent $\gamma D\to\phi D$, and incoherent $\gamma D\to\phi np$
  reactions aimed at (i) studying effect of elimination of the
  isovector $\pi$ meson exchange in coherent $\gamma D$ reaction, and
  (ii) extracting  observables for the $\gamma n$ reaction.
  The elimination of the $\pi$ meson exchange has two
  consequences. One is the relative decrease of channels
  with spin-conserving amplitudes, which result in an increase of
  the relative contributions of
  the spin flip transitions. This leads to an enhancement of the
  corresponding spin density matrix elements.
  Another one is related to a strong suppression of the amplitude
  with un-natural parity exchange and shift $\rho^1_{1-1}$ matrix
  element towards 0.5.
  We got a common description of $\phi$ meson decay
  distributions for  $\gamma p\to\phi p$ and
  incoherent $\gamma D\to\phi np$ reactions confirming
  the reliability of our model for the $\gamma n$ reaction.

  To summarize we can conclude, that the existing
  experimental data (including also very recent data) on  $\gamma p$,
  coherent $\gamma D\to\phi D$, and incoherent $\gamma D\to\phi np$
  reactions in the diffraction region at low energies support the model
  based on the dominance of
  the Donnachie-Landshoff Pomeron plus $\pi, \eta$ exchange
  channels with a relatively
  weak $\eta$ meson contribution.
  For a definite conclusion about a possible manifestation of exotic
  channels one has to improve the resolution of the data with
  providing additional information on the channels with
  spin- and double-spin flip transitions being  sensitive
  to properties of the photoproduction amplitude in $\gamma p$
  and $\gamma D$ reaction. This problem
  may be studied experimentally at the electron and photon
  facilities at LEPS of SPring-8, JLab, Crystal-Barrel of ELSA, and
  GRAAL of ESRF.

 \acknowledgments

 We thank
 W.C.~Chang, S.~Dat\'e, H.~Ejiri,  M.~Fujiwara, T.~Mibe,
 T.~Nakano, and Y.~Ohashi for many fruitful discussions and comments.
 One of the authors (A.I.T.) appreciates  colleagues in FZD for the
 hospitality.
 This work was supported by BMBF grant 06DR136 and GSI-FE.



\begin{thebibliography}{30} 

\bibitem{NakanoToki}
T.~Nakano and H.~Toki, in {\it Proceedings of the International
Workshop  on Exiting Physics and New Accelerator Facilities},
SPring-8, Hyogo, 1997 (World Scientific Singapore, 1998), p. 48.


\bibitem{PL97}
 M.A.~Pichowsky and T.-S.~H.~Lee, Phys. Rev. D
 {\bf 56}, 1644 (1997).

\bibitem{TOYM98}
 A.I.~Titov, Y.~Oh, S.N.~Yang, and T.~Morii,
 Phys.\ Rev.\ C {\bf 58}, 2429 (1998).

\bibitem{ZLB98}
Q.~Zhao, Z.~Li, and C.~Bennhold,
  Phys. Lett. B {\bf 436}, 42 (1998);
  Phys. Rev. C {\bf 58}, 2393 (1998).

\bibitem{Will98}
R. A. Williams,
  Phys. Rev. C {\bf 57}, 223 (1998).

\bibitem{Laget2000}
J.-M.~Laget,
  Phys. Lett. B {\bf 489}, 313 (2000).


\bibitem{TL03}
 A. Titov and T.-S.H. Lee,
Phys. Rev. C {\bf 67}, 065205 (2003).

\bibitem{DL84-92}
A.~Donnachie and P.\,V.~Landshoff,
  Phys.\ Lett.\  \textbf{B185}, 403 (1987);
  Nucl.\ Phys.\ \textbf{B244}, 322 (1984);
  \textit{ibid.\/} \textbf{B267}, 690 (1986).

\bibitem{Ryskin93}
  M.~G.~Ryskin,
  Z.\ Phys.\  C {\bf 57}, 89 (1993).

\bibitem{Cudell97}
  J.~R.~Cudell and I.~Royen,
  Phys.\ Lett.\  B {\bf 397}, 317 (1997).

\bibitem{Mibe05}
 T.~Mibe {\it et al}
 [LEPS Collaboration],
 Phys.\ Rev.\ Lett. C {\bf 95}, 182001 (2005).

\bibitem{FKS97}
 L.L.~Frankfurt, J.~Mutzbauer, W.~Koepf, G. Piller,
 M.~Sargsian, and M.I.~Strikman,
 Nucl. Phys. A {\bf 622} 511 (1997);
L.~Frankfurt, G.~Piller, M.~Sargsian, and M.~Strikman, Eur.\
Phys.\ J.\ A {\bf 2}, 301 (1998).

\bibitem{Rogers06}
  T.~C.~Rogers, M.~M.~Sargsian, and M.~I.~Strikman,
  Phys.\ Rev.\  C {\bf 73}, 045202 (2006).

\bibitem{TFL02}
  A.~I.~Titov, M.~Fujiwara and T.~S.~H.~Lee,
  Phys.\ Rev.\  C {\bf 66}, 022202 (2002).

 \bibitem{WenChen07}
 W.~C.~Chang {\it et al}
 [LEPS Collaboration],
 nucl-ex/0703034


\bibitem{Mibe07}
 T.~Mibe {\it et al}
 [CLAS Collaboration],
 nucl-ex/0703013

\bibitem{Landshoff87}
  P.~V.~Landshoff and O.~Nachtmann,
  Z.\ Phys.\  C {\bf 35}, 405 (1987).

\bibitem{QCDSR}
  S.L.~Zhu,
  Phys. Rev. C {\bf 61}, 065205 (2000).


\bibitem{XPT}
  J.~Piekarewicz,
  Phys. Rev. C {\bf 48}, 1535 (1993).
\bibitem{etaphoto}

\bibitem{etaphoto}
  L.~Tiator, C.~Bennhold and S.~S.~Kamalov,
  Nucl.\ Phys.\  A {\bf 580}, 455 (1994);
  M.~Kirchbach and L.~Tiator,
  Nucl.\ Phys.\  A {\bf 604}, 385 (1996).

\bibitem{TLTS99}
A.I. Titov, T.-S.~H. Lee, H.~Toki, and O.~Streltsova,
  Phys. Rev. C {\bf 60}, 035205 (1999).

\bibitem{Paris}
 M.~Lacombe, B.~Loiseau, R.~Vinh Mau, J.~Cote, P.~Pires,
 and R.~de~Tourreil, Phys. Lett. B {\bf 101}, 139 (1981).

\bibitem{ParisD}
M.~Lacombe, B.~Loiseau, J.M.~Richard, R.~Vinh Mau, J.~Cote,
P.~Pires and R.~de~Tourreil,
Phys.\ Rev.\ C {\bf 21}, 861 (1980).

\bibitem{SAPHIR}
  J.~Barth {\it et al.},
  Eur.\ Phys.\ J.\  A {\bf 17}, 269 (2003).

\bibitem{Bonn}
H.J.~Besch et al., Nucl. Phys. B {\bf 70}, 257 (1974).

\bibitem{JLab36}
  E.~Anciant {\it et al.}  [CLAS Collaboration],
  Phys.\ Rev.\ Lett.\  {\bf 85}, 4682 (2000).

\bibitem{OldData}
The Durham Data Base. HEP REACTION DATA.

\bibitem{WenChen07add}
 We appraciate W.C. Chang for providing us
 $[\gamma D,\phi]$ missing mass distribution
 at $\Delta_t=0-0.1$~GeV$^2$, measured
 by the LEPS Collaboration.

\bibitem{Schilling70}
 K.~Schilling, P.~Seyboth, and G.~E.~Wolf,
  Nucl.\ Phys.\  B {\bf 15}, 397 (1970)
  [Erratum-ibid.\  B {\bf 18}, 332 (1970)].

\bibitem{Mibe05add}
 We appraciate T.~Mibe for providing us
 the experimental data of the
 angular distribution $W^L(\Phi-\Phi)$ for
 the $\gamma p\to \phi p$ reaction measured
 by the LEPS Collaboration.
\end{thebibliography}
 \end{document}